\documentclass[]{aa}   
\usepackage{graphicx}
\usepackage{multirow}
\usepackage{subcaption}
\usepackage[normalem]{ulem}

\usepackage{color}

\definecolor{tom}{rgb}{0.0, 0., 1.0}
\definecolor{vaibhav}{rgb}{0.1, 0.5, 0.2}

\definecolor{lianne}{rgb}{0.7, 0.0, 0.5}

\makeatletter
\newcommand*{\rom}[1]{\expandafter\@slowromancap\romannumeral #1@}
\makeatother

\usepackage{siunitx}
\newcommand{\Alfvenic}{Alfv\'enic }
\newcommand{\Alfven}{Alfv\'en }
\newcommand{\TDC}{\tau_{\text{DC}} }
\newcommand{\TAC}{\tau_{\text{AC}} }
\usepackage[utf8]{inputenc}
\usepackage{fourier} 
\usepackage{array}
\usepackage{makecell}

\usepackage{placeins}

\begin{document}
\title{Forward modelling of heating within a coronal arcade}
\author{L.~E. Fyfe \inst{1} \and T.~A. Howson \inst{1} \and I. De Moortel \inst{1, 2}}
\institute{School of Mathematics and Statistics, University of St. Andrews, St. Andrews, Fife, KY16 9SS, U.K. \and Rosseland Centre for Solar Physics, University of Oslo, PO Box 1029  Blindern, NO-0315 Oslo, Norway}

\abstract{}
{We investigate the synthetic observational signatures from numerical models of coronal heating in an arcade to determine what features are associated with such heating, and what tools can be used to identify them.}
{We consider two simulations of coronal arcades driven by footpoint motions with different characteristic timescales. Forward modelling is then used, and the synthetic emission data is analysed.}
{The total intensity and Doppler velocities clearly show the magnetic structure of the coronal arcade. Contrasts in the local Doppler shift also highlight the locations of separatrix surfaces. The distinguishing feature of the AC and DC models is that of the frequencies. Through FFT analysis of the Doppler velocities, when short timescale footpoint motions are present, higher frequencies are observed. For longer timescale motions, the dominant signal is that of lower frequencies; however, higher frequencies were also detected, which matched the natural Alfv\'en frequency of the background magnetic field. Alfv\'enic wave signatures were identified in both models with fast wave signatures observable in the AC model. Finally, the estimates of the kinetic energy using the Doppler shifts were found to significantly underestimate within these models.}
{Observables identified within this article were from features such as Alfv\'en waves, fast waves, the arcade structure and separatrix surfaces.  The two models were differentiated by examining the frequencies present. The Doppler velocities cannot provide accurate estimates of the total kinetic energy or the component parallel to the LOS. This is due to some plasma outside the formation temperature range of the ion, the multi-directional driver, and cancellation of the velocity along the LOS. The impact each factor has on the estimation is dependent on the set up of the model and the chosen emission line.}


\maketitle


\section{Introduction}\label{sec:introduction}

It is well established that the solar corona is heated up to millions of degrees. In order to investigate the source of this heating, various models have been suggested which could generate such high temperatures. Coronal heating theories can be found in reviews by, for example \citet{WalshIreland2003,Klimchuk2006, Klimchuk2015,ParnellDeMoortel2012, DeMoortelBrowning2015}. Convective motions at the Sun's surface, are thought to be the source of energy required to heat the solar corona. Energy flux is injected into the solar atmosphere due to these photospheric motions, and is subsequently converted into heat. However, how and where this energy is released is still unclear.


 The models proposed for releasing energy in the solar corona largely fall into two broad groups; AC heating and DC heating. These two categories differ according to the timescale of the photospheric motions that drive them. When timescales are short (in comparison to the \Alfven travel time along a coronal loop), we refer to AC heating. On the other hand, if the driving timescale is long, we are in the DC heating regime. In either case, in order to obtain any significant energy release, small scales need to be generated within the magnetic and/or velocity fields. For comprehensive coronal heating reviews we direct the reader to \citep[][]{ParnellDeMoortel2012,Arregui2015,DeMoortelBrowning2015,Klimchuk2015,VanDoorsselaereNakariakov2020,VanDoorsselaereSrivastava2020}.

Even though there has been an enhancement to the spatial and temporal resolution in imaging and spectroscopic instruments over the past couple of decades, observing coronal heating is still a challenge. One technique used to further the understanding and knowledge around the AC and DC heating mechanisms is to compare observations to synthetic spectroscopic data obtained from numerical simulations of energy release. For example, \citet{KlimchukKarpen2010} use the one-dimensional time-dependent hydrodynamic equations to model loops in thermal nonequilibrium and subsequently determine if the synthetic EUV observables of these warm loops satisfy the five observational characteristics summarised by the authors. \citet{LionelleWinebarger2013} proceed to expand this work by modelling fully 3D simulations which take the geometry of the coronal magnetic field into account, as well as applying a more realistic heating mechanism. \citet{LionelleWinebarger2013} found that the signatures from their model matched the observed properties of EUV loops. Studies have also been undertaken to distinguish the observational signatures of different heating profiles (e.g. impulsive vs steady heating)\citep[e.g.][]{Cargill1994, WinebargerLionello2018}.


MHD waves and oscillations are known to be ubiquitous within the solar atmosphere. Not only are they thought to contribute to the heating of the solar corona but waves also provide a crucial tool for indirectly measuring atmospheric conditions through coronal seismology \citep[for example, see the review by][]{DeMoortelNakariakov2012}. By analysing observational signals of MHD waves and oscillations, estimates for the plasma parameters can be found. Often these are difficult to determine with direct observations/measurements. Increasingly, authors are using synthetic observables generated by numerical models to understand the potential and limitations of coronal seismology. \citep[e.g.][]{AntolinVanDoorsselaere2013,YuanVanDoorsselaere2015, YuanVanDoorsselaere2016}.



In this study, we investigate such synthetic spectroscopic data. The two numerical simulation for which the synthetic observables are generated from is that of a coronal arcade \citep{HowsonDeMoortelFyfe2020} where the  characteristic timescale of the driver imposed at the base of the corona, is different between the two simulations. This will help further understand potential observables from similar heating within a coronal arcade. In particular, we search for specific diagnostics which will allow us to distinguish between the two mechanisms. In Sect. \ref{sect_numerical_model}, we give a brief overview of the numerical models and results from \citet{HowsonDeMoortelFyfe2020}. In Sect. \ref{fomo_section}, the synthetic observables are analysed by investigating the imaging and spectral signatures. Finally, Sect. \ref{sec_Discussion} summarises and concludes the findings of this study.

\section{Numerical models} \label{sect_numerical_model}

\subsection{Setup}

To begin, we provide a brief description of the numerical model from which we generate the synthetic emission data and subsequently discuss within this article (see Sect. \ref{fomo_section}). For additional information on the analysis of this simulation we direct the reader to \citet{HowsonDeMoortelFyfe2020}. 

To investigate the effects driving timescales have on heating within a coronal arcade, \citet{HowsonDeMoortelFyfe2020} perform three simulations, each with different driving timescales. These are also subdivided into ideal, resistive and viscous cases. The numerical simulations are implemented using the Lagrangian-remap code, Lare3d \citep{ArberLongbottom2001}, which solves the non-ideal 3D MHD equations in normalised form. In \citet{HowsonDeMoortelFyfe2020}, gravity, thermal conduction and radiative losses are neglected from their model. However, resistivity and viscosity are included within separate simulations, as well as including an ideal case. Within the resistive regime a step function is implemented such that the resistivity equals zero strictly below $z = 1$ Mm and equals $\eta_0$ for $z \ge 1$ Mm, where $\eta_0$ results in a characteristic magnetic Reynolds number of approximately $10^4$. This is calculated using the numerical resolution and typical length scales and velocities within the simulations. By implementing zero resistivity on the lower $z$ boundary, there is limited (only numerical) magnetic field line slippage through the velocity field. As for the viscosity, it is uniform through space and time and generates a characteristic fluid Reynolds number of approximately $10^3$. Again, this value is calculated for the numerical resolution and typical length scales and velocities in these simulations.

\begin{figure}[t!]
\centering
\includegraphics[width=0.4\textwidth]{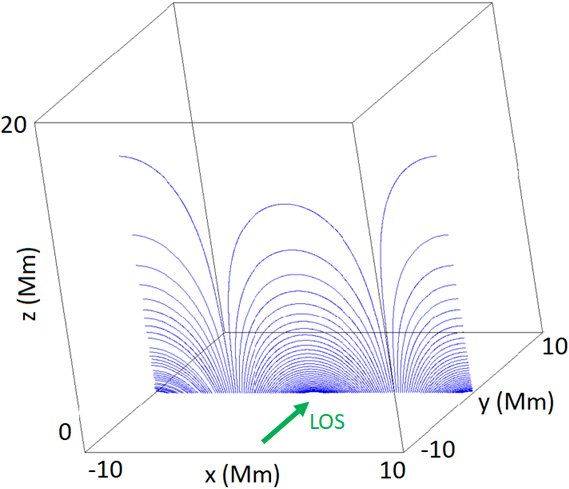}
\caption{Illustration of the potential arcade which is invariant along the $y$ axis. Green arrow depicts the LOS used during the forward modelling (see Sect. \ref{fomo_section}). This figure is modified from \citet{HowsonDeMoortelFyfe2020}.}
\label{mag_field_arcade}
\end{figure}

The coronal arcade is initially contained within a homogeneous plasma with a temperature and density of approximately 1 MK and 1.67 $\times10^{-12}\text{ kg}\text{ m}^{-3}$, respectively. The magnetic field has the form $\vec{B}(x,z) = \left(B_x,0,B_z\right)$ where

\begin{eqnarray}
    \ B_x(x, z) &=& B_0\text{ cos}\left(\frac{\pi x}{L}\right)\exp{\left(\frac{-\pi z}{L}\right)},\nonumber \; \\
    \ B_z(x, z) &=& -B_0\text{ sin}\left(\frac{\pi x}{L}\right)\exp{\left(\frac{-\pi z}{L}\right)},   \;
\end{eqnarray}

\noindent and with $B_0 = 100$ G and $L = 10$ Mm. This magnetic field is a potential field which is invariant along the $y$ axis (see Fig. \ref{mag_field_arcade}). The numerical domain is comprised of $256^3$ grid cells with physical dimensions of  $-10 \text{ Mm}\leq x,y \leq  10 \text{ Mm}$ and $0 \text{ Mm}\leq z \leq  20 \text{ Mm}$.

\citet{HowsonDeMoortelFyfe2020} implement random footpoint motions on the base of the domain (assumed to be representative of photospheric motions). More specifically, the driver is composed of the summation of 2D Gaussians (in space and in time) and has the form $\vec{v} = \left(v_x, v_y, 0\right)$ where

\begin{eqnarray}
    \ v_x &=& \sum_{i=1}^{N} v_i\cos\left(\theta_i\right)\exp\Bigg\{\frac{-\left(r-r_i\right)^2}{l_i^2}\Bigg\}\exp\Bigg\{\frac{-\left(t-t_i\right)^2}{\tau_i^2}\Bigg\}, \nonumber \; \\
    \  v_y &=& \sum_{i=1}^{N} v_i\sin\left(\theta_i\right)\exp\Bigg\{\frac{-\left(r-r_i\right)^2}{l_i^2}\Bigg\}\exp\Bigg\{\frac{-\left(t-t_i\right)^2}{\tau_i^2}\Bigg\}.  \;
\end{eqnarray}

\noindent For each $i$, $v_i$, $\theta_i$, $r_i$ and $l_i$ are the peak amplitude, direction, centre and length scale of the Gaussian components and $t_i$ and $\tau_i$ represent the time of peak amplitude and the duration of the individual Gaussian components, respectively. These variables are derived from the following statistical distributions,

\begin{eqnarray}
\  v_i \sim \mathcal{N}\left(v_{\mu}, \frac{v_{\mu}^2}{25}\right), \quad \theta_i \sim \mathcal{U}\left(0, 2\pi\right),\quad r_i\sim\mathcal{U}\left(-L,L\right),\quad \; \nonumber \\
\ l_i \sim\mathcal{N}\left(\frac{L}{4},\frac{L^2}{400}\right), \quad t_i \sim\mathcal{U}\left(t_s, t_f\right), \quad\tau_i \sim \mathcal{N}\left(\tau_{\mu}, \frac{\tau_{\mu}^2}{16}\right),\; 
\end{eqnarray}

\noindent where $\mathcal{N} \left(\mu, \sigma^2\right)$ and $\mathcal{U}\left(u_1, u_2\right)$ are the normal and uniform distributions, respectively. The mean and variance of the normal distribution are $\mu$ and $\sigma^2$, respectively and $u_1$ and $u_2$ denote the lower and upper bounds for the uniform distribution. 

Three different driving timescales are analysed in \citet{HowsonDeMoortelFyfe2020}, of which we shall consider the two extremes, namely  $\tau_{\mu}$ = 15 s and $\tau_{\mu}$ = 300 s. The shorter and longer timescales are chosen to replicate the AC and DC heating mechanisms, respectively and hence we shall denote them by $\TAC$ and $\TDC$, respectively. The spatio-temporal average of the velocities is set to $1.2 \text{ km}\text{ s}^{-1}$ in both cases by selecting an appropriate value for $v_{\mu}$. Finally, $N$ is chosen to be a function of the timescale $\tau_{\mu}$ to ensure that a similar number of components in the summation are active at any given time. 

The boundaries within this model are periodic on the $x$ and $y$ boundaries. The gradients of the variables are set to zero on both $z$ boundaries with the exception of the velocity driver which is imposed on the lower boundary. Near the top of the domain, a damping layer is implemented to prevent the reflection of upward waves back into the simulation. It is constructed using $\vec{v}:= a\left(z\right)\vec{v}$, where $a$ decreases linearly from $a(18 \text{ Mm}) = 1$ to $a(20 \text{ Mm}) = 0.99$.

\subsection{Evolution}

Before interpreting the synthetic emission data produced by the forward modelling (see Sect. \ref{fomo_section}), an understanding of the system's evolution is helpful. More specifically, we analyse the behaviour of the density, temperature and velocity field.

\begin{figure}[ht!]
\centering
\vspace{0cm}
\begin{subfigure}{0.5\textwidth}
  \centering
  \makebox[0pt]{\includegraphics[width=0.89\textwidth]{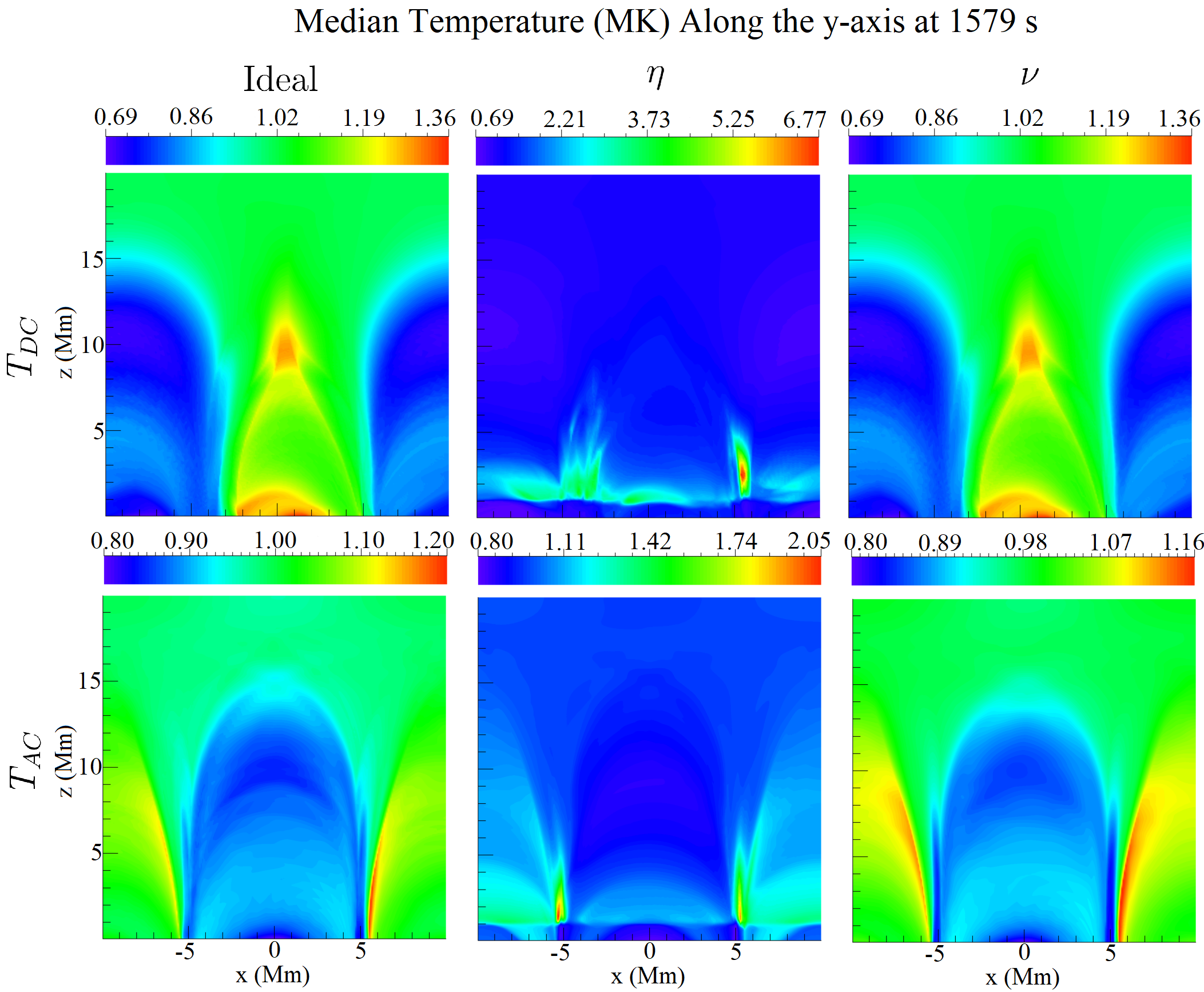}}
  \caption{}
  \label{med_temp}
\end{subfigure}
\begin{subfigure}{0.5\textwidth}
  \centering
  \hspace{0cm}
\makebox[0pt]{\includegraphics[width=0.89\textwidth]{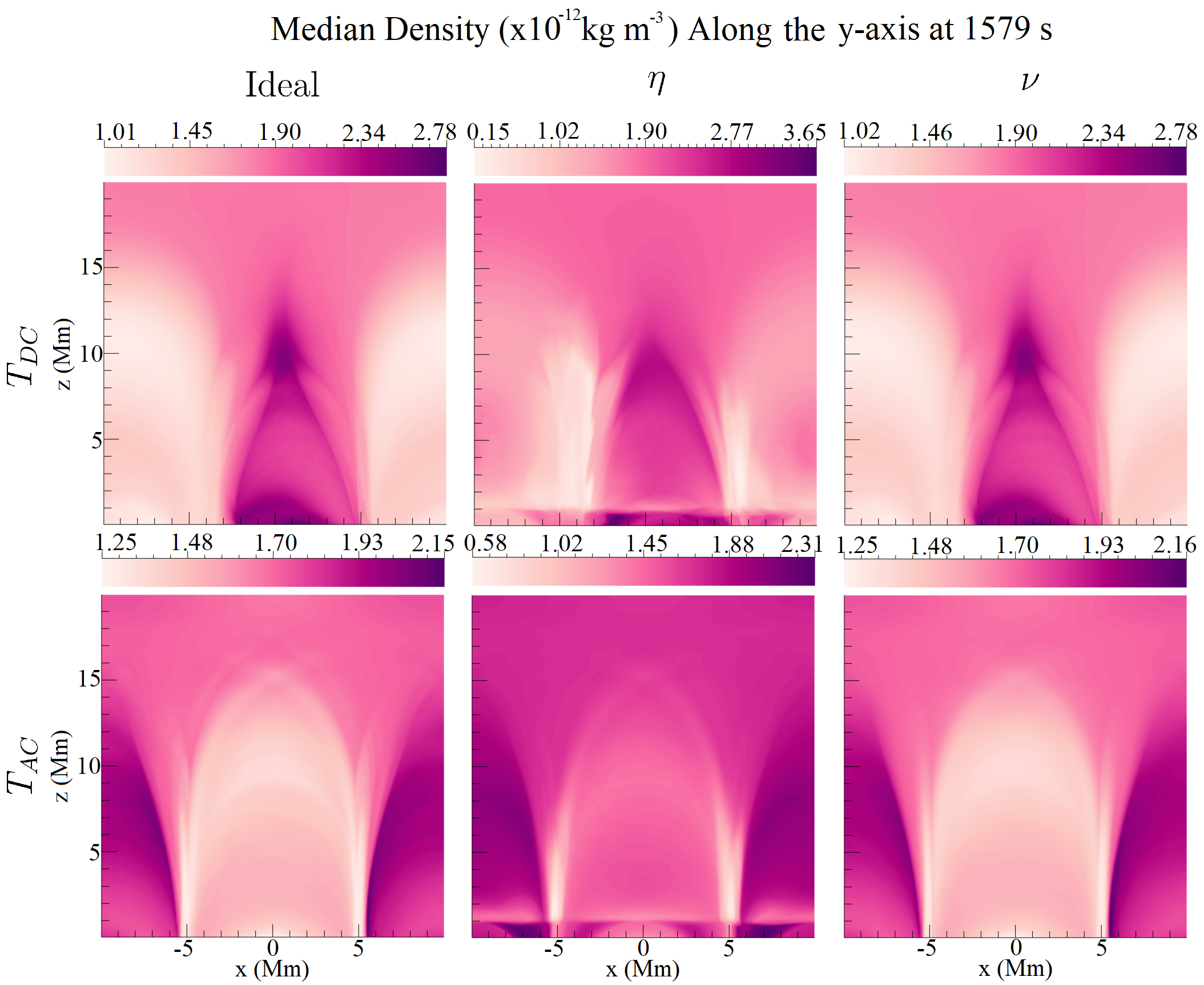}}
  \caption{}
  \label{med_rho}
\end{subfigure}
\begin{subfigure}{0.5\textwidth}
  \centering
  \hspace{0cm}
\makebox[0pt]{\includegraphics[width=0.89\textwidth]{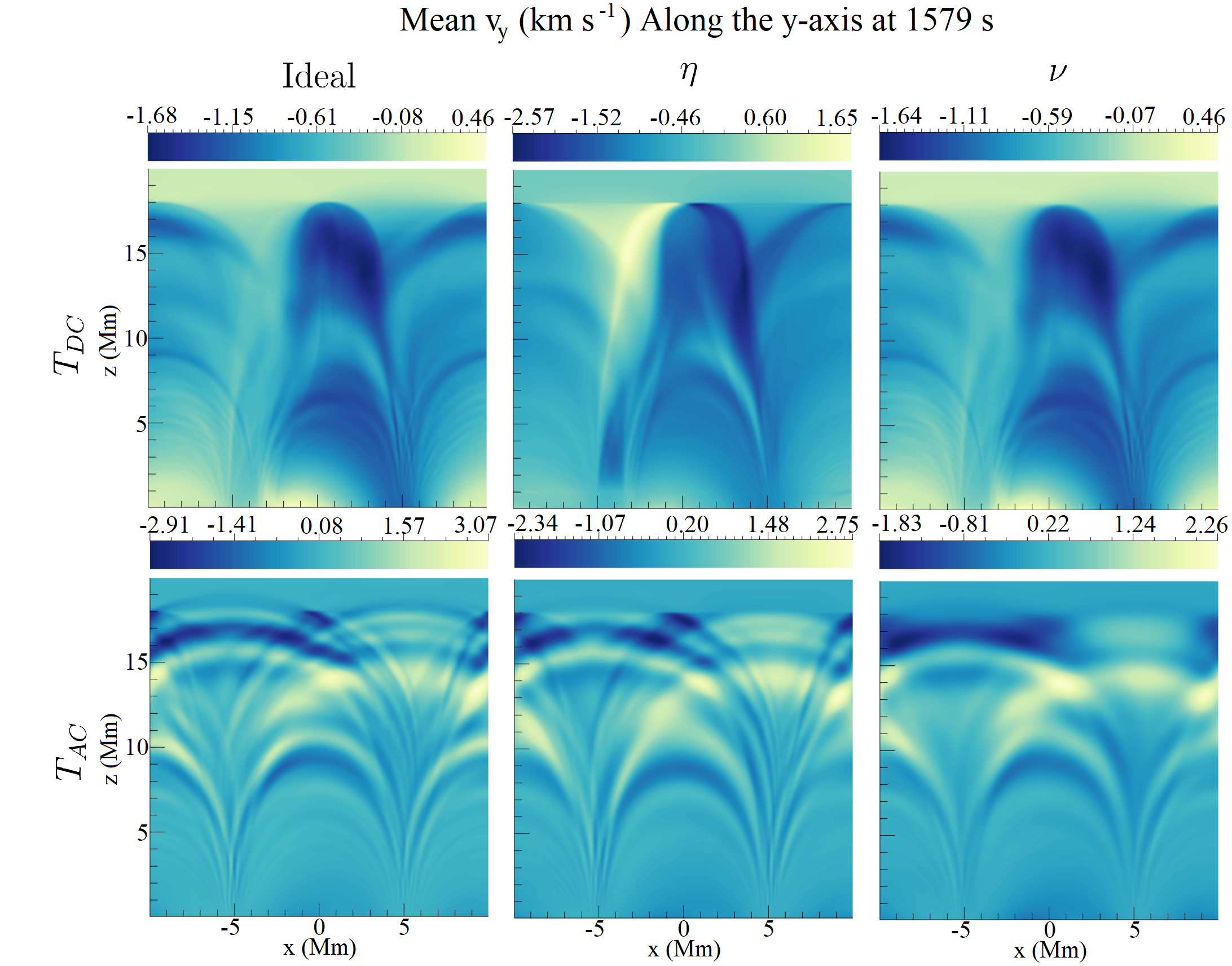}}
  \caption{}
  \label{med_rho}
\end{subfigure}
\caption{Median temperature (a), density (b) and average $v_y$ along the $y$ axis at 1579 s in the ideal (column 1), resistive (column 2) and viscous (column 3) simulations using the $\TDC$ (row 1) and $\TAC$ (row 2) boundary drivers.}
\label{med_temp_rho_mean_vel}
\end{figure}

Figure \ref{med_temp_rho_mean_vel} illustrates the temperature (Fig. 2a), density (Fig. 2b) and velocity profiles (Fig. 2c) at $t = 1579$ s by taking the median temperature and density (as the mean profile is skewed by large temperatures due to the exclusion of thermal conduction) and average $v_y$ along the $y$ axis (invariant direction). This instance in time is chosen as it is fairly representative of the features seen during the complete simulation. We show the results of the ideal (first column), resistive (second column) and viscous (third column) simulations. Within all the simulations, these plasma variable profiles reveal the coronal arcade's magnetic structure. Due to the absence of thermal conduction, the temperature profiles in the resistive regime do not show these structures as clearly as it potentially would have in coronal conditions. With thermal conduction, these structures would be enhanced due to the diffusion of thermal energy along the magnetic field lines. As for the high density structures in the simulated coronal loops, these may not be realistic as gravity is ignored in these simulations. For both the temperature and density profiles of the resistive simulations, the effect of the zero resistivity region is present for $z<1$ Mm. Within the velocity profiles, not only do we see arcade structures, but we observe a signature of the drivers. It is evident that the frequent and short-lived nature of the $\TAC$ driver (lower panels of Fig. 2c) generates more distinct small scale spatial features than in the $\TDC$ simulations (upper panels).

\citet{HowsonDeMoortelFyfe2020} found that greater plasma heating occurs when the longer timescale driver was implemented ($\TDC$) in the resistive regime where magnetic energy can be released. The DC-like velocity perturbations are associated with a larger Poynting flux (on average) compared to the $\TAC$ driver. In Fig. \ref{med_temp}, in the resistive regime (second column), larger temperatures are present. As optically thin radiation, thermal conduction and mass exchange with the lower atmospheric layers are not modelled here, a large increase in the temperature is observed.


The largest magnetic field perturbations are found near the base of the coronal domain and, as a result, large currents tend to form here. In addition, we also see that large gradients in the magnetic field develop near the separatrix surfaces at $x = \pm 5$ Mm. As the magnetic connectivity changes across these planes, footpoint motions on neighbouring field lines can generate strong currents and ultimately, Ohmic heating \citep{HowsonDeMoortelFyfe2020}. This is apparent in the second column of Fig. \ref{med_temp}, where temperature enhancements are present at the base of the separatrix surfaces. In addition to this Ohmic heating, the dissipation of energy via viscous effects also causes heating. \citet{HowsonDeMoortelFyfe2020} found that the largest velocity gradients form between the coronal arcades. This was measured using the vorticity of the velocity field. Due to the short timescale driver in the $\TAC$ simulation, higher frequency velocity perturbations are present which subsequently generate shorter wave lengths and hence, larger velocity gradients. This results in more viscous heating within the viscous $\TAC$ simulation. This viscous heating results in the higher temperature plasma observed on the separatrix surfaces ($x \approx \pm $ 5) in the bottom right panel of Fig.\ref{med_temp}.

\citet{HowsonDeMoortelFyfe2020} found that the kinetic energy is significantly less than the perturbed magnetic energy during all the simulations; hence, the dissipation of kinetic energy within the viscous simulations is expected to have less of an impact than the magnetic energy which is dissipated within the resistive regime. Therefore, Ohmic heating is expected to dominate the plasma heating over the viscous heating (for similar fluid and magnetic Reynolds numbers). Indeed, within Fig. \ref{med_temp}, the heating observed in the resistive simulation (second column) contains hotter plasma than that in the viscous simulation (third column). 

\section{Forward modelling} \label{fomo_section}

\subsection{Emission lines and LOS}
To forward model the numerical simulations the FoMo code is used \citep{VanDoorsselaereAntolin2016}. This code generates optically thin EUV and UV emission lines using the CHIANTI atomic database \citep{DereLandi1997,LandiYoung2013}.

For the duration of the simulation, the median temperature of the plasma ranges from approximately 1 MK to 1.5 MK. Nonetheless, the mean temperature can vary from approximately 1 MK to 10 MK. This is due to a small number of locations within the domain of the $\TDC$ resistive simulation in which the temperatures are unrealistically high. Approximately 5\% of the plasma in this $\TDC$ simulation is greater than 6.31 MK (i.e. approximately the maximum temperature that Fe \rom{16} detects). These localised larger temperatures are partially due to the exclusion of optically thin radiation and thermal conduction from the model. In order to cover the majority of the temperatures present, the Fe \rom{9} 171\AA, Fe \rom{12} 193 $\AA\text{ }$ and Fe \rom{16} 335$\AA\text{ }$ emission lines are used. Their peak formation temperatures are approximately log($T$) = 5.92 ($\sim 0.839$ MK), log($T$) = 6.20 ($\sim 1.57$ MK) and log($T$) = 6.42 ($\sim 2.65$ MK), respectively.


Finally, we chose a line-of-sight (LOS) angle aligned with the invariant direction of the initial magnetic field (i.e. the $y$ axis). This is illustrated by the green arrow in Fig.\ref{mag_field_arcade}.

\subsection{Imaging signatures}


\begin{figure}[t!]
\centering
\vspace{0cm}
\begin{subfigure}{0.5\textwidth}
  \centering
  \makebox[0pt]{\includegraphics[width=0.94\textwidth]{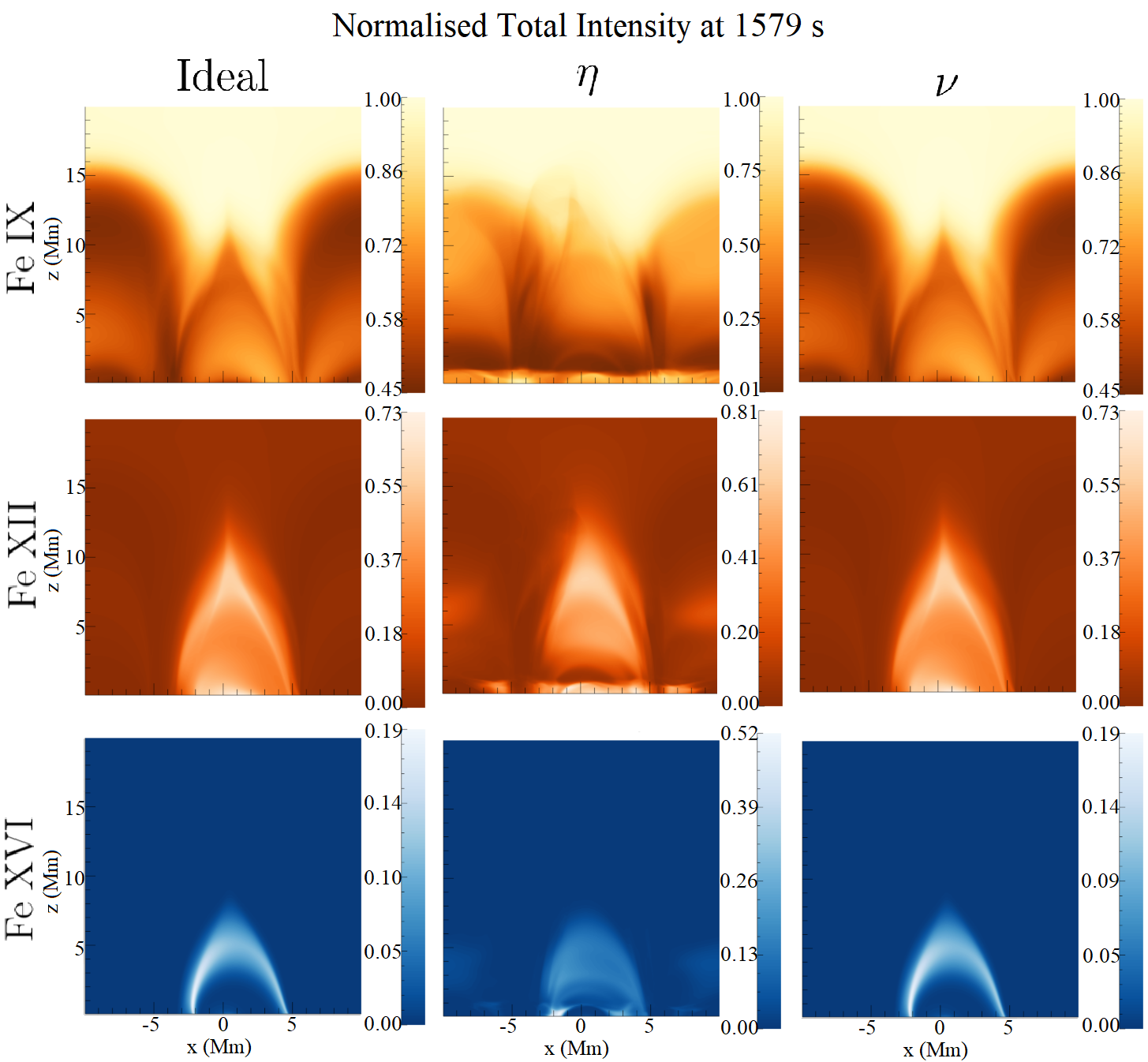}}
  \caption{}
  \label{braid_tot_int}
\end{subfigure}
\begin{subfigure}{0.5\textwidth}
  \centering
  \hspace{0cm}
\makebox[0pt]{\includegraphics[width=0.94\textwidth]{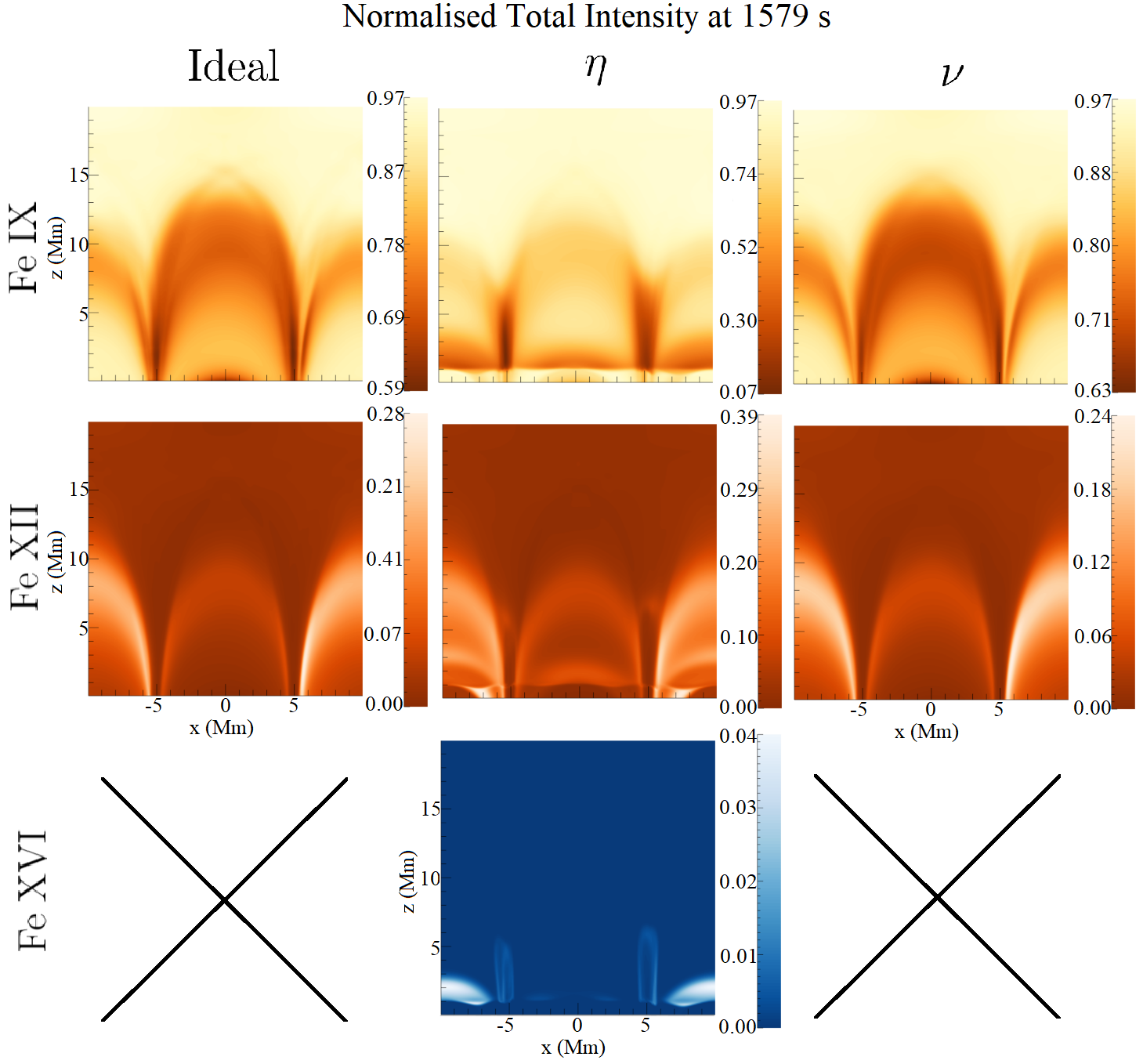}}
  \caption{}
  \label{wave_tot_int}
\end{subfigure}
\caption{Normalised synthetic total intensity at 1579 s in Fe \rom{9} (row 1), Fe \rom{12} (row 2) and Fe \rom{16} (row 3) for the ideal (column 1), resistive (column 2) and viscous (column 3) regimes using the (a) $\TDC$ and (b) $\TAC$ boundary drivers. Every panel is normalised by the spatio-temporal maximum of all simulations in each emission line. The bottom left and right panels in (b) are neglected as very little is detected in the Fe \rom{16} line.} 
\label{tot_int}
\end{figure}

The (synthetic) total intensity at $t = 1579 \text{ s}$ in all three emission lines during the ideal, resistive and viscous simulations with the (a) $\TDC$ and (b) $\TAC$ drivers, is illustrated in Fig. \ref{tot_int}. As expected, given the temperature and density profiles (see Fig. \ref{med_temp_rho_mean_vel}), the magnetic structure of the coronal arcade is apparent within the total intensity, with the exception of the $\TAC$ simulations in Fe \rom{16} (see below). The visible structure for $z<1$ Mm in the resistive simulations, caused by the zero resistivity region, is also observable in the total intensity. In general, Fe \rom{9} detects the cooler plasma while Fe \rom{12} observes the hotter plasma. This hotter plasma is typically found within the central dense arcade in the $\TDC$ simulations and along the longer loops bound by the separatrix surfaces in the $\TAC$ simulations, in the locations of highest vorticity. 

In this study, the $\TDC$ simulation generates hotter plasma than the $\TAC$ simulation. Indeed, when comparing the two driving mechanisms, the slow timescale driving generates greater maximum synthetic emission in the plane-of-the-sky (POS) than the AC-like driver for the Fe \rom{12} line. 

\begin{figure}[b!]
\centering
\vspace{0cm}
\begin{subfigure}{0.5\textwidth}
  \centering
  \makebox[0pt]{\includegraphics[width=0.94\textwidth]{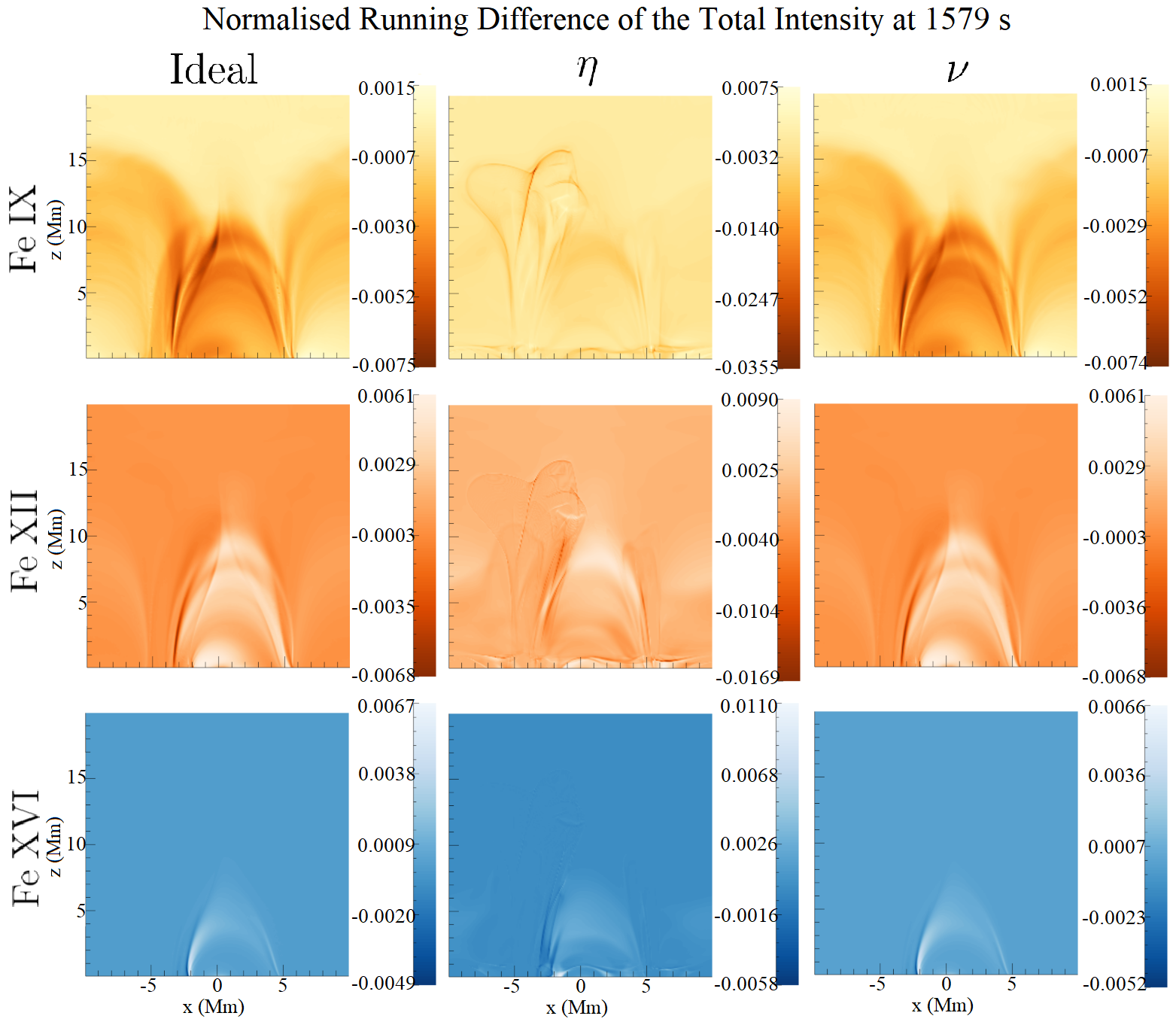}}
  \caption{}
  \label{braid_rundiff}
\end{subfigure}
\begin{subfigure}{0.5\textwidth}
  \centering
  \hspace{0cm}
\makebox[0pt]{\includegraphics[width=0.94\textwidth]{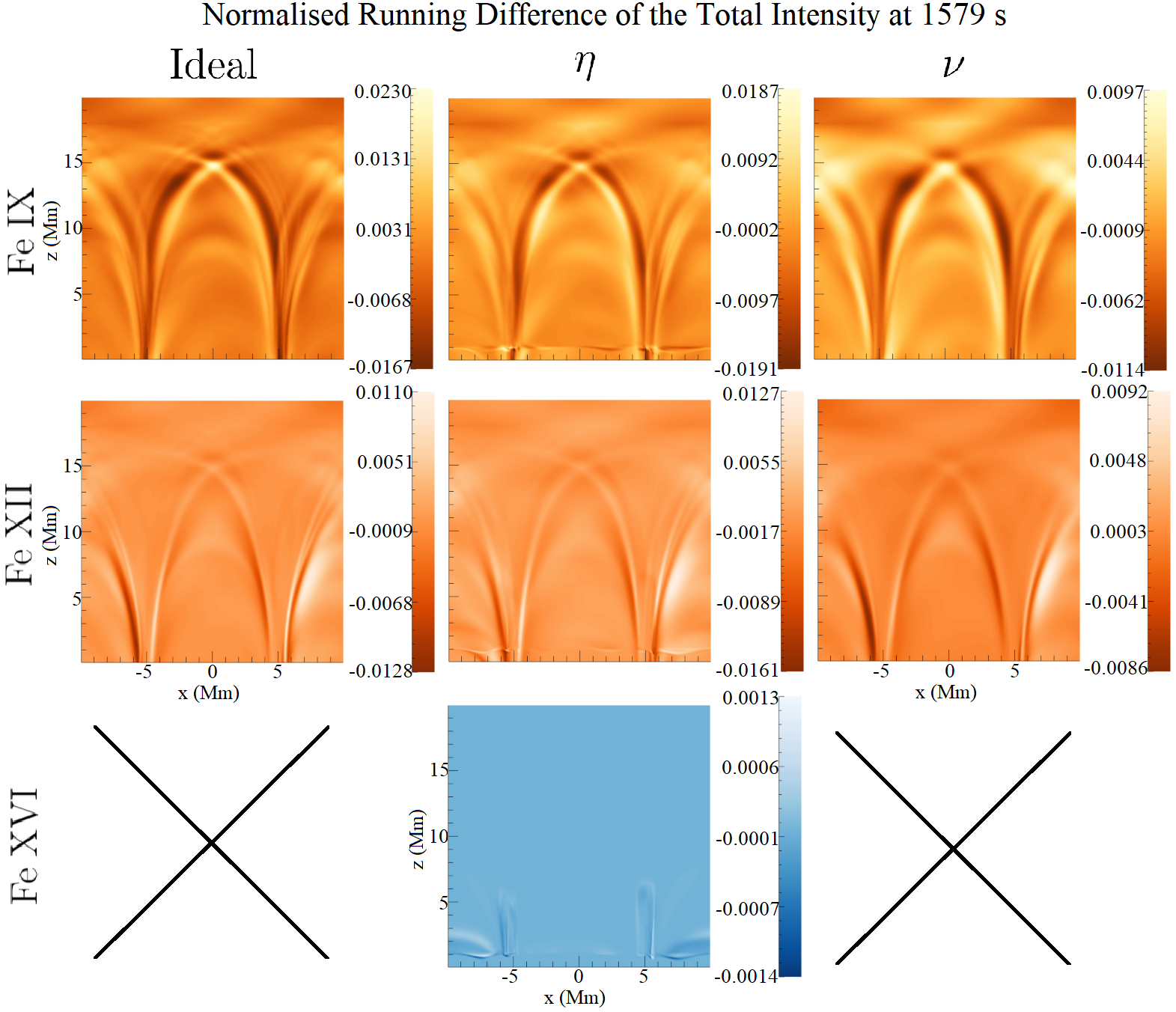}}
  \caption{}
  \label{wave_rundiff}
\end{subfigure}
\caption{Running difference (cadence of 7.2 s) of the normalised synthetic total intensity at 1579 s in Fe \rom{9} (row 1), Fe \rom{12} (row 2) and Fe \rom{16} (row 3) for the ideal (column 1), resistive (column 2) and viscous (column 3) simulations using the (a) $\TDC$ and (b) $\TAC$ boundary drivers. Normalisation and neglected panels are as described in Figure \ref{tot_int}} 
\label{rundiff}
\end{figure}

As with the Fe \rom{12} synthetic emission, the Fe \rom{16} emission line detects the hotter plasma found within the central arcade in the $\TDC$  simulation and in regions of low lying magnetic field line in the $\TAC$  simulation. However, due to the reduced heating present in the $\TAC$ simulation (in comparison to the $\TDC$ simulation) the plasma is not hot enough to be detected by the Fe \rom{16} line within the ideal and viscous regimes. Even within the resistive simulation, very little extremely hot plasma is detected at this time.


The arcade structures are also apparent in the running difference of the synthetic total intensity (current time frame minus the previous with a cadence of 7.2 s) as seen in Fig. \ref{rundiff}. Again, the only exception to this is the Fe \rom{16} emission in the $\TAC$ simulations. Finer scale features are present within the running difference, with the shorter timescale driver ($\TAC$) creating smaller spatial features as the driver is providing higher frequency velocity perturbations to the system than those in the $\TDC$ simulations. 

Only within the intensity running difference of the resistive $\TDC$ simulation is a structure seen rising between the left and central arcades (see the top two panel of the second column in Fig. \ref{braid_rundiff} where the top of the structure has reached $z\approx 15$ Mm). This is the result of adiabatic expansion cause by large temperatures from at the base, between the two arcades, as a result of Ohmic heating. This feature may not be present if thermal conduction was included in the model as it would decrease the large field-aligned temperatures present in our current simulation.




\subsection{Doppler velocities} \label{spec_sig_sect}

\begin{figure}[t!]
\centering
\vspace{0cm}
\begin{subfigure}{0.5\textwidth}
  \centering
  \makebox[0pt]{\includegraphics[width=0.94\textwidth]{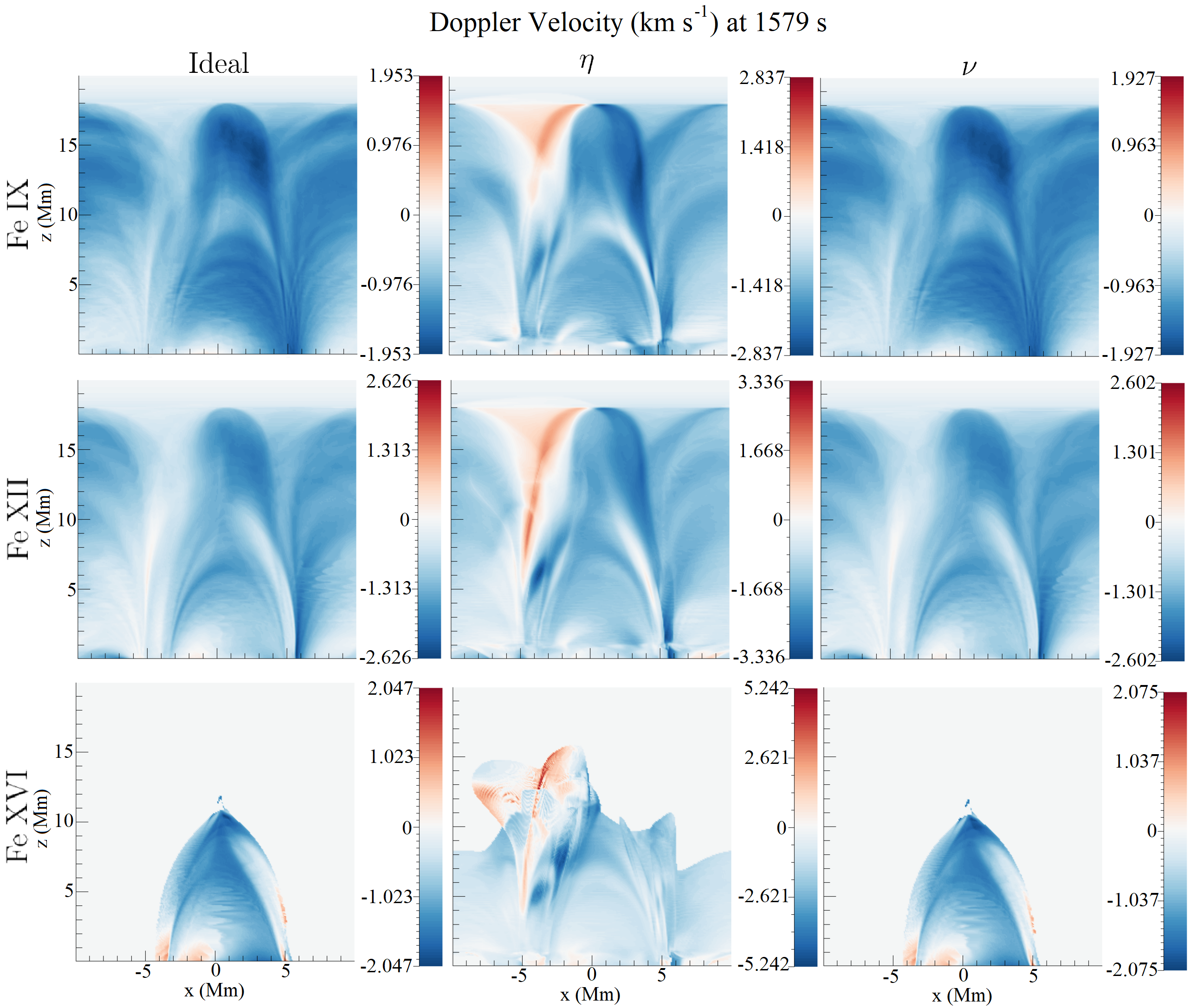}}
  \caption{}
  \label{braid_dv}
\end{subfigure}
\begin{subfigure}{0.5\textwidth}
  \centering
  \hspace{0cm}
\makebox[0pt]{\includegraphics[width=0.94\textwidth]{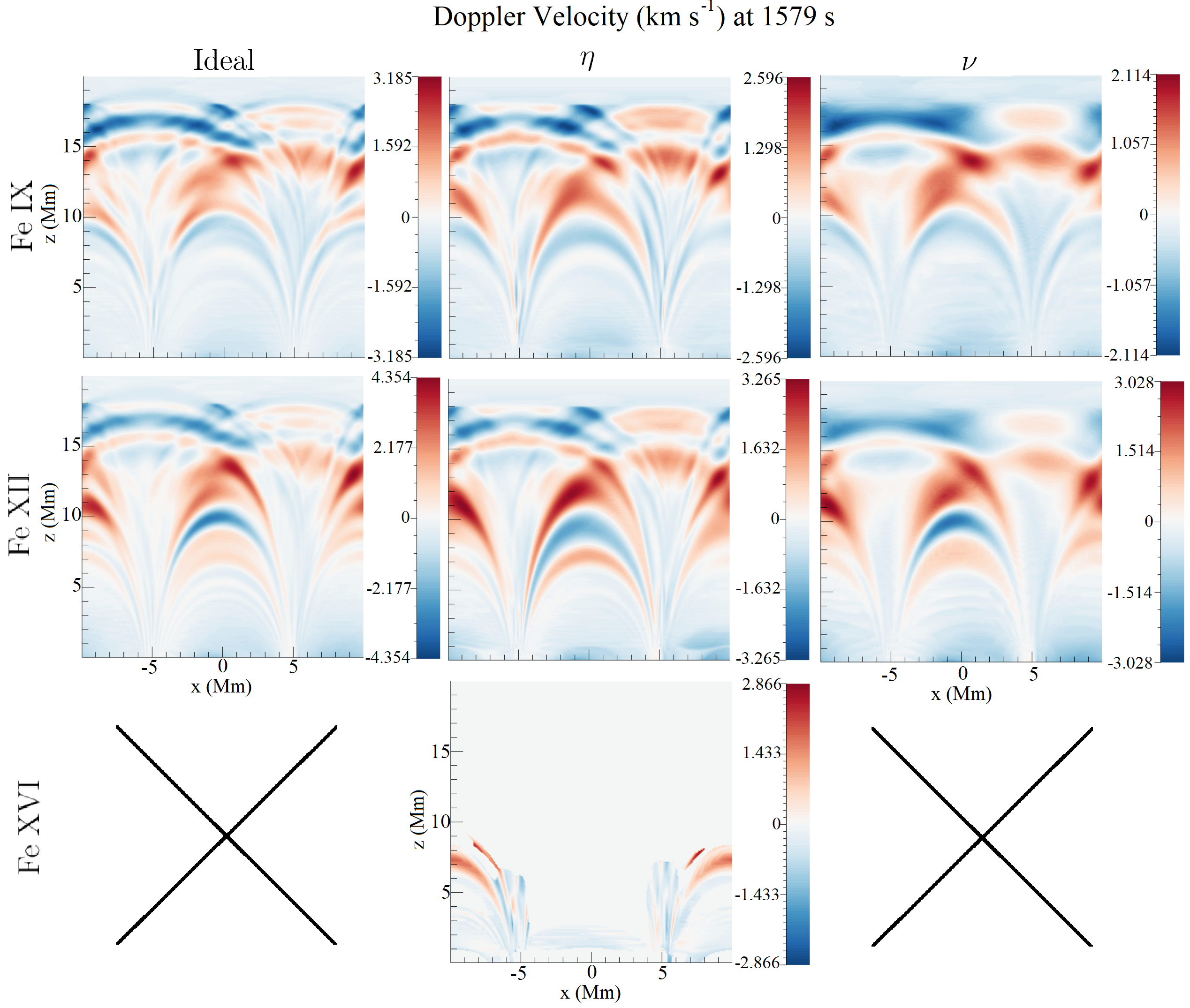}}
  \caption{}
  \label{wave_dv}
\end{subfigure}
\caption{Doppler velocity at 1579 s in Fe \rom{9} (row 1), Fe \rom{12} (row 2) and Fe \rom{16} (row 3) for the ideal (column 1), resistive (column 2) and viscous (column 3) simulations using the (a) $\TDC$ and (b) $\TAC$ boundary drivers.} 
\label{dv_contour}
\end{figure}

The smaller spatial scale features observed during the total intensity running difference of the $\TAC$ simulations are also present within the Doppler velocity signatures. The Doppler velocities are calculated by fitting a (single) Gaussian to the specific intensity and measuring the shift of the Gaussian's peak from the rest wavelength of the emission line. Figure \ref{dv_contour} illustrates the Doppler velocity at 1579 s in the (a) $\TDC$ and (b) $\TAC$ simulations, where the rows represent the different emission lines and the columns denote the ideal and non-ideal regimes. At this time, there is a dominant blue shift throughout the POS within the $\TDC$ simulation. This low variability in the Doppler velocity is due to the long timescale velocity perturbations. Similarly, at other times a dominating red shift is present and the Doppler shift average to zero over the duration of the simulation. Conversely, the $\TAC$ driver produces altering red-blue Doppler shifts with smaller spatial features. The arcade's structure is also observable within both simulation, although it is clearer in the $\TAC$ simulation. During the simulations, the general structure and spatial features observed in the Doppler velocity profiles do not change significantly. The magnitude and direction of the Doppler shifts do indeed change but do not alter the analysis provided within this article.

\begin{figure*}[ht!]
\centering
\vspace{0cm}
\begin{subfigure}{0.8\textwidth}
  \centering
  \makebox[0pt]{\includegraphics[width=1.\textwidth]{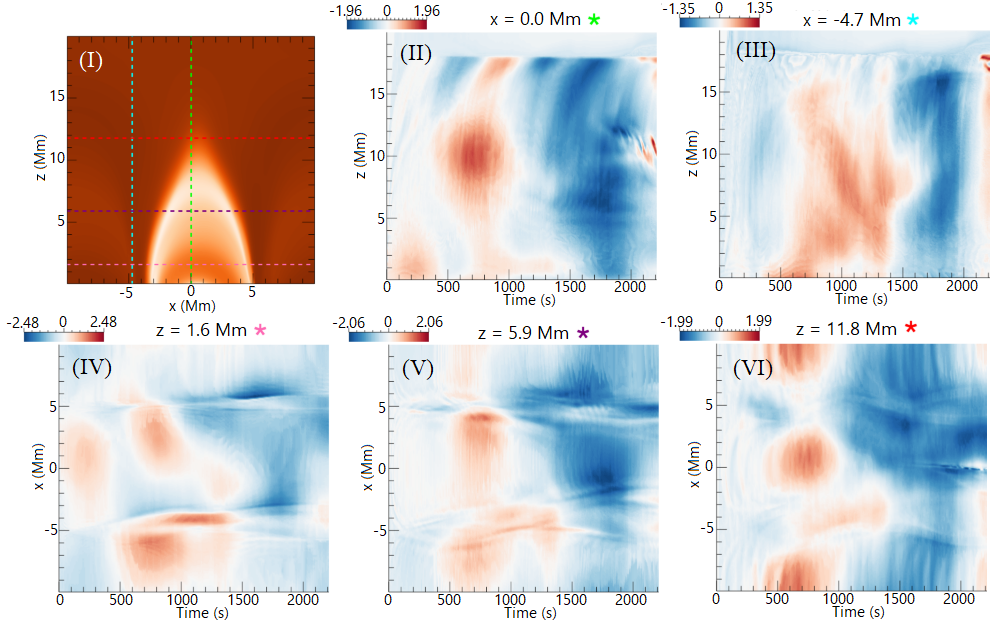}}
  \caption{}
  \label{braid_dv_slit}
\end{subfigure}
\begin{subfigure}{0.8\textwidth}
  \centering
  \hspace{0cm}
\makebox[0pt]{\includegraphics[width=1.\textwidth]{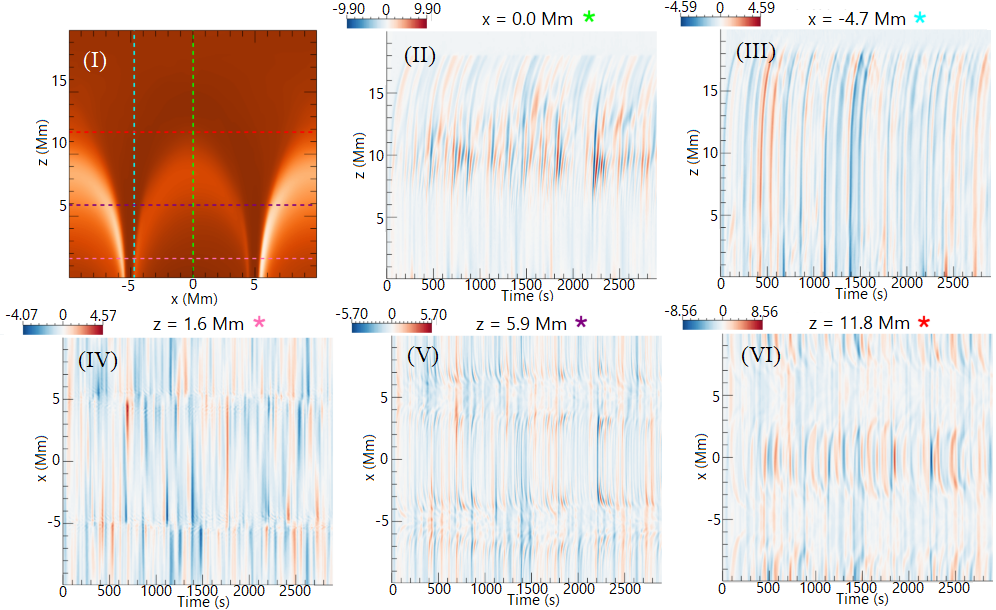}}
  \caption{}
  \label{wave_dv_slit}
\end{subfigure}
\caption{Doppler velocity for Fe \rom{12} along various slits as a function of time in the ideal (a) $\TDC$ and (b) $\TAC$ simulations in $\text{km}\text{ s}^{-1}$ in panels (\rom{2})-(\rom{6}). The slits are colour coded in the synthetic total intensity figure in panel (\rom{1}).}
\label{dv_slits}
\end{figure*}

It is somewhat apparent from the Doppler velocity profiles that higher frequencies features are present when shorter timescale footpoint motions are present (i.e. the $\TAC$ simulations compared to $\TDC$ simulations). This is confirmed through Fig. \ref{dv_slits}, which illustrates the Fe \rom{12} Doppler velocity as a function of time along five different slits across the POS for the ideal (a) $\TDC$ and (b) $\TAC$ simulations. The location of these slits in relation to the arcade are indicated in the top left panel. From this point onward, any analysis shall only be applied to the ideal regime in Fe \rom{12} (unless stated otherwise) as there is no significant difference between that and the other emission lines and non-ideal simulations.

\begin{figure}[t!]
\centering
\includegraphics[width=0.5\textwidth]{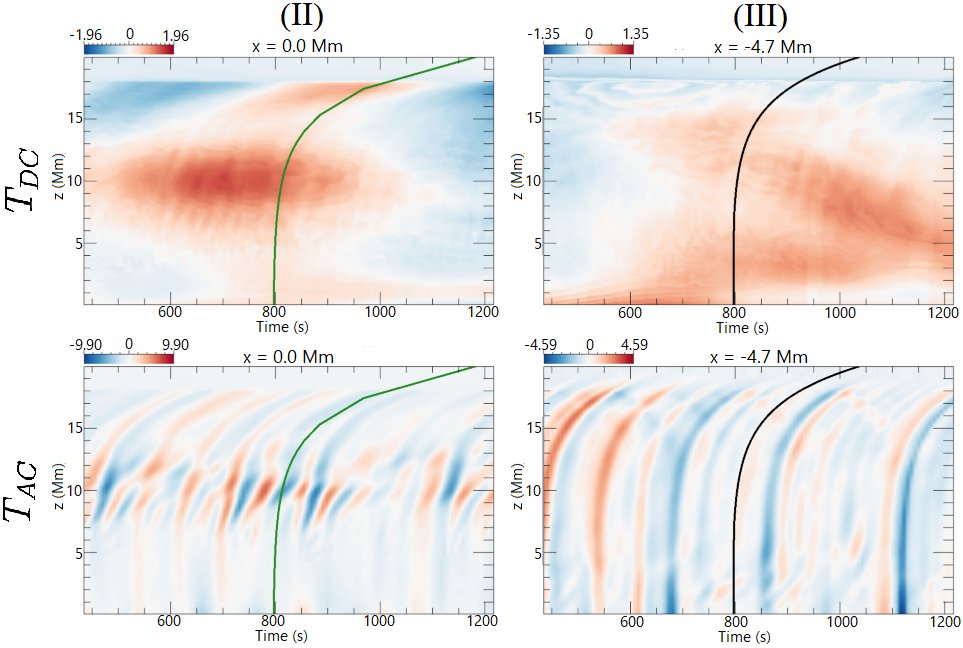}
\caption{As panels \rom{2} (through the apex) and \rom{3} (between arcades) in Fig. \ref{dv_slits} for the $\TDC$ (top row) and $\TAC$ (bottom row) simulations with a truncated time axis between approximately 450 s and 1200 s.  The solid lines are the expected signatures of an \Alfven wave excited on the lower boundary at approximately $t = 800$ s. These are calculated along a single magnetic field line (black: right hand panels) and taking multiple magnetic field lines into account (green: left hand panels).}
\label{dv_slit_zoom_plus_va}
\end{figure}

\begin{figure}[t!]
\centering
\vspace{0cm}
\begin{subfigure}{0.24\textwidth}
  \centering
  \makebox[0pt]{\includegraphics[width=1.\textwidth]{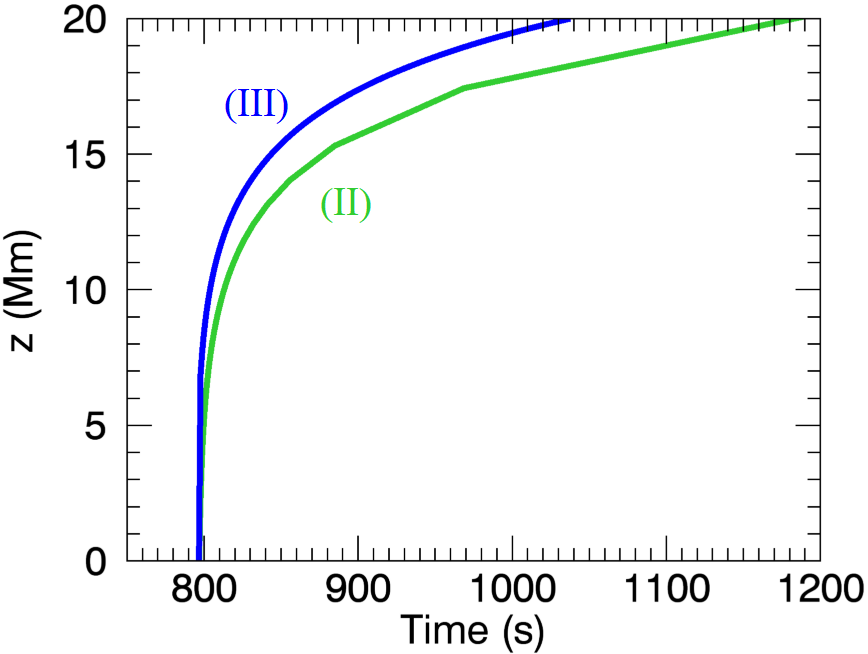}}
  \caption{}
  \label{vA_apex_and_between}
\end{subfigure}%
\begin{subfigure}{0.26\textwidth}
  \centering
  \hspace{0cm}
  \makebox[0pt]{\includegraphics[width=1.\textwidth]{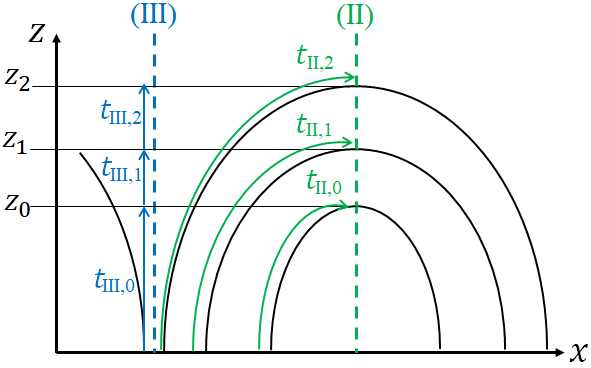}}
  \caption{}
  \label{vA_distance_explained}
\end{subfigure}
\caption{As the lines seen in Fig. \ref{dv_slit_zoom_plus_va} now illustrated in panel (a) along with (b) an illustration of how they are calculated. The black lines depict the arcades structure. For each height $z_{i}$, $t_{\text{\rom{2}},i}$ and $t_{\text{\rom{3}},i}$ denote the time it takes from $z = 0$ to reach that height on slits (\rom{2}: green dashed line) and (\rom{3}: blue dashed line), respectively.}
\label{va_slits}
\end{figure}

In both Figs. \ref{braid_dv_slit} and \ref{wave_dv_slit}, propagating features are observed travelling from lower to higher altitudes along the two vertical slits (see panels \rom{2}:through the apex of the arcade and \rom{3}:between the arcade structures). To examine these four panels in more detail Fig. \ref{dv_slit_zoom_plus_va} was constructed, which restricts the time axis for the four mentioned panels. The solid lines starting on the bottom $z$ boundary at approximately 800 s illustrate how a disturbance travelling at the local \Alfven speed would appear. In particular, we assumed that the lower boundary was perturbed by a linear transverse wave at this time and calculated when it would reach different heights along the slit. These two solid lines (green: column \rom{2} and black: column \rom{3}) are calculated slightly differently. When observing between the arcades (column \rom{3}), the black line is attained by approximately following one single magnetic field line as the slit and field line will have better alignment compared to the green line which passes through the apex of the arcade (column \rom{2}). The green line is therefore calculated by taking multiple magnetic field lines into account. This means the \Alfven travel time is obtained from multiple footpoints to the position in which those field lines meet the slit.  Figure \ref{vA_apex_and_between} illustrates the signature that we would expect to see along the two slits (blue: between arcades and green: through the apex of the arcade) if a linear transverse wave was generated at 800 s on the bottom boundary. These are the same lines as in Fig. \ref{dv_slit_zoom_plus_va} with the black line now blue in Fig. \ref{vA_apex_and_between}. A pictorial representation of the field lines travelled by the wave is given in Fig. \ref{vA_distance_explained} before they reach the slits.

From Fig. \ref{vA_apex_and_between}, we would expect to see shallower gradients when observations along the slit which crosses through the apex -- compared to between the arcades -- are made. This is the result of the increased amount of time it takes for \Alfvenic waves to travel to higher altitudes due to the longer magnetic field lines which cross the slit through the arcade's apex. Indeed, the propagating structures seen in Fig. \ref{dv_slit_zoom_plus_va} (and the zoomed out panels \rom{2} and \rom{3} in Fig. \ref{dv_slits}) coincide with the shape of the overplotted lines. Finally, below the velocity damping layer ($z= 18$ Mm) (where no observable motion is present), the velocity of the propagating \Alfvenic wave features is seen to decrease with altitude. This is simply due to the decrease in \Alfven speed with increasing height ($z$) \citep{HowsonDeMoortelFyfe2020}.We note that the change in gradient with altitude corresponds to the decrease in the magnetic field strength (and hence the local \Alfven speed).

\begin{figure}[t!]
\centering
\includegraphics[width=0.45\textwidth]{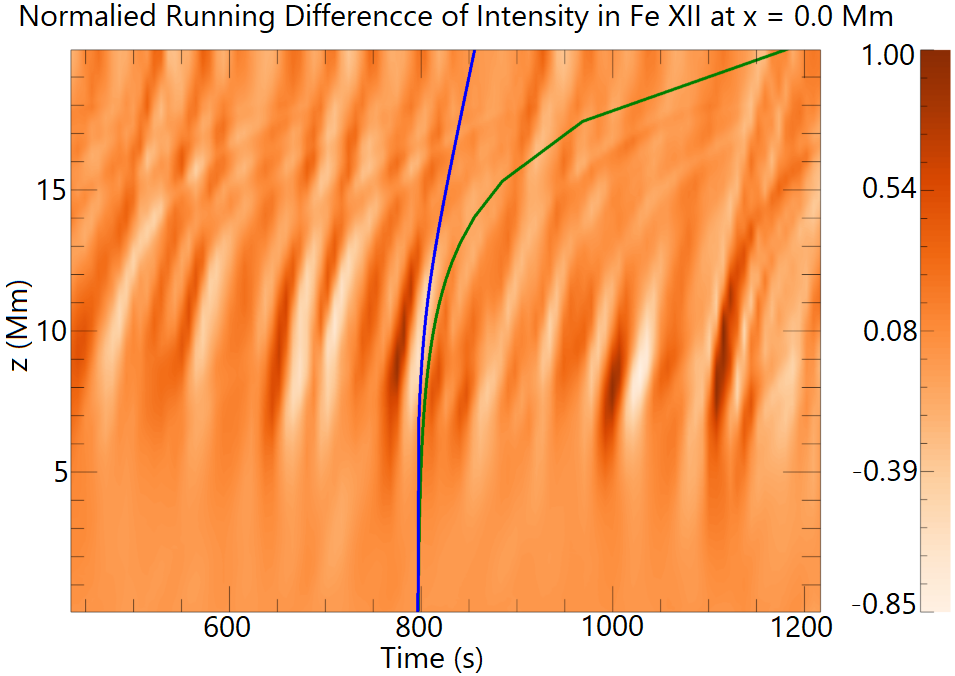}
\caption{Temporal evolution of the normalised running difference of the Fe \rom{12} synthetic total intensity along $x = 0$ Mm in the ideal $\TAC$ simulation. The blue and green lines represent the distance travelled by a fast and \Alfven wave, respectively.}
\label{rundiff_with_va_vf}
\end{figure}

Signatures of the \Alfvenic waves are also observed in the running difference of the total intensity; however, they are only detected for the $\TAC$ (not the $\TDC$) case. Figure \ref{rundiff_with_va_vf} illustrates the evolution of the normalised running difference of the Fe \rom{12} total intensity along the slit $x = 0$ Mm (through the arcade's apex) in the ideal regime (this corresponds to the green slit in panel \rom{1} of Fig. \ref{wave_dv_slit} and the Doppler velocity time-distance contour in the bottom left hand panel of Fig. \ref{dv_slit_zoom_plus_va}). The green line in Fig. \ref{rundiff_with_va_vf} (same as the green line in the bottom left panel of Fig. \ref{dv_slit_zoom_plus_va}), represents the signature of an \Alfven wave excited at approximately $t= 800$ s. It is seen to match the shallower gradient features present in the running difference.
 
In Fig. \ref{rundiff_with_va_vf}, there are also features with larger gradients caused by fast waves. The signature of a fast wave, which is excited at approximately $t = 800$ s from the base of the domain, is given by the blue line and coincides with the intensity running difference. Similar to the \Alfvenic wave, the fast wave slows with increasing altitude due to the decreasing fast speed \citep{HowsonDeMoortelFyfe2020}. The fast wave is observable in the intensity running difference as it is compressible; however, evidence of the fast wave is not seen in the previous Doppler velocity analysis (see bottom left panel of Fig. \ref{dv_slit_zoom_plus_va}) as the signature of \Alfvenic wave dominates the signal.

As we are able to observe the signatures of the fast and \Alfvenic waves, it is possible to estimate the local fast and \Alfven speeds. Using seismology one can estimate, for example, the plasma temperature or the magnetic field strength. However, in our simulation, the uncertainty on the wave speed estimates leads to a wide range of temperature estimates, which does not provide a better constraint on the plasma temperature than the temperature formation range of our emission lines. 

Now we shall return to Fig. \ref{dv_slits} and consider the lower two horizontal slits, namely panels \rom{4} and \rom{5}. Within the $\TDC$ simulation a sharp contrast in the local Doppler velocity is present along $x \approx \pm 5$. This coincides with the location of the separatrix surfaces where larger velocity gradients are predominately found.
The reason for the asymmetry of the flows on either side of the separatrix surface is that the magnetic connectivity changes across this plane and thus neighbouring field lines may not be excited in the same way \citep{HowsonDeMoortelFyfe2020}. This contrast in the Doppler velocity is also present, but less obvious, in the $\TAC$ simulations due to the shorter temporal features.

\begin{figure}[h!]
\centering
\vspace{0cm}
\begin{subfigure}{0.25\textwidth}
  \centering
  \hspace{0cm}
  \makebox[0pt]{\includegraphics[width=1.\textwidth]{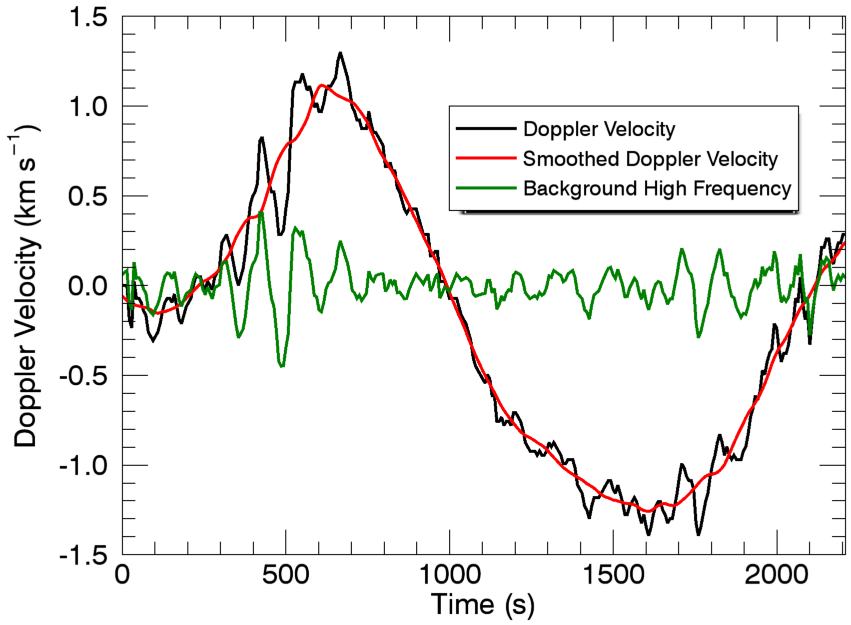}}
  \caption{}
  \label{dv_with_smooth}
\end{subfigure}%
\begin{subfigure}{0.25\textwidth}
  \centering
  \makebox[0pt]{\includegraphics[width=1.\textwidth]{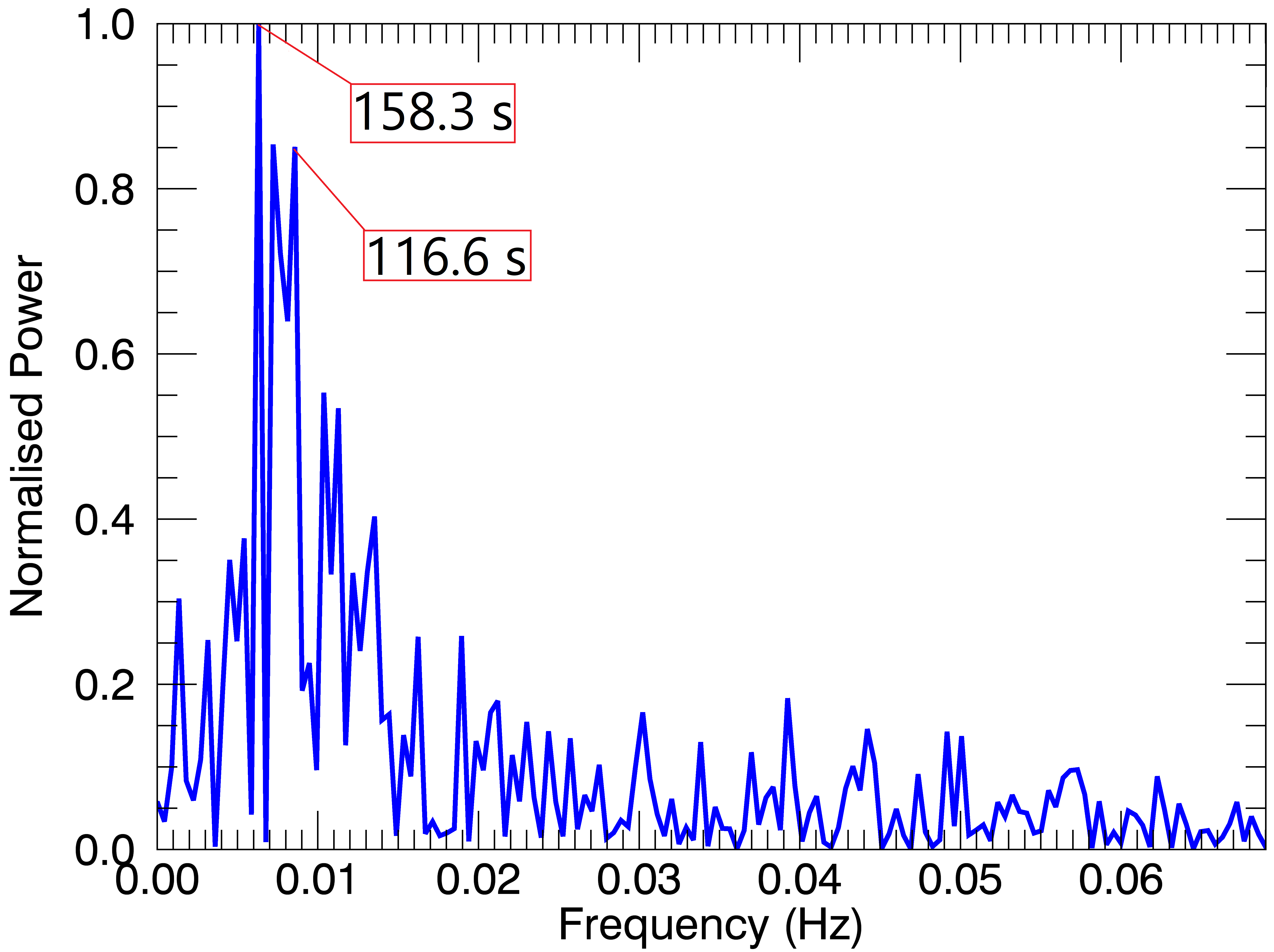}}
  \caption{}
  \label{fft_dv_smooth}
\end{subfigure}
\caption{Panel (a) contains the Doppler velocity (black) as a function of time at $(x, z) = (-9.9, 11.8)$ Mm, along with a smoothed version of the Doppler velocity curve (red) and the difference between the Doppler velocity and the smoothed Doppler velocity (green). Panel (b) contains the corresponding FFT of the high frequency signal (green curve in panel (a)).}
\label{background_frequency}
\end{figure}

For the $\TDC$ simulations (see Fig. \ref{braid_dv_slit}), the Doppler velocity in the horizontal slits (see panels \rom{4}-\rom{6}) shows higher frequency perturbations as well as large spatial structures. To analyse these higher frequencies further, a point in the POS, namely $(x,z)=(-9.9., 11.8)$ Mm, is taken and the behaviour of the Doppler velocity is plotted (black line) in Fig. \ref{dv_with_smooth}. The longer period is evidently dominating the signal and originates from the DC boundary driver. To analyse the higher frequency (green line), a smoothed version of the longer period (red line) is subtracted from the original signal. An FFT (using the standard \textit{IDL} routine) is then applied to the high frequency signal and shown in Fig. \ref{fft_dv_smooth}, where the dominant wave periods are annotated within the red boxes. An estimate for the period of the fundamental mode along the initial magnetic field line which passes through the point $(x,z)=(-9.9,11.8)$ Mm (corresponds to panel \rom{6} of Fig. \ref{braid_dv_slit}) is obtained by calculating the \Alfven travel time $\left(\tau_{\text{A}}\right)$ using 

\begin{equation}
   \tau_{\text{A}} = \int_{L}^{} \frac{ds}{v_{\text{A}}(s)}, 
\end{equation}

\noindent where $ds$ is an infinitesimal length along the magnetic field line and $v_{\text{A}}$ is the local \Alfven speed. The initial conditions are used for this calculation. The period of the fundamental mode is found to be 116 s. This is approximately equal to the period of the high frequency Doppler velocity signal seen in Fig. \ref{background_frequency}; hence it is reasonable to assume that the high frequency Doppler velocity signal is due to the excitation of standing \Alfvenic waves.

\begin{figure*}[t!]
\centering
\vspace{0cm}
\begin{subfigure}{0.8\textwidth}
  \centering
  \makebox[0pt]{\includegraphics[width=1.\textwidth]{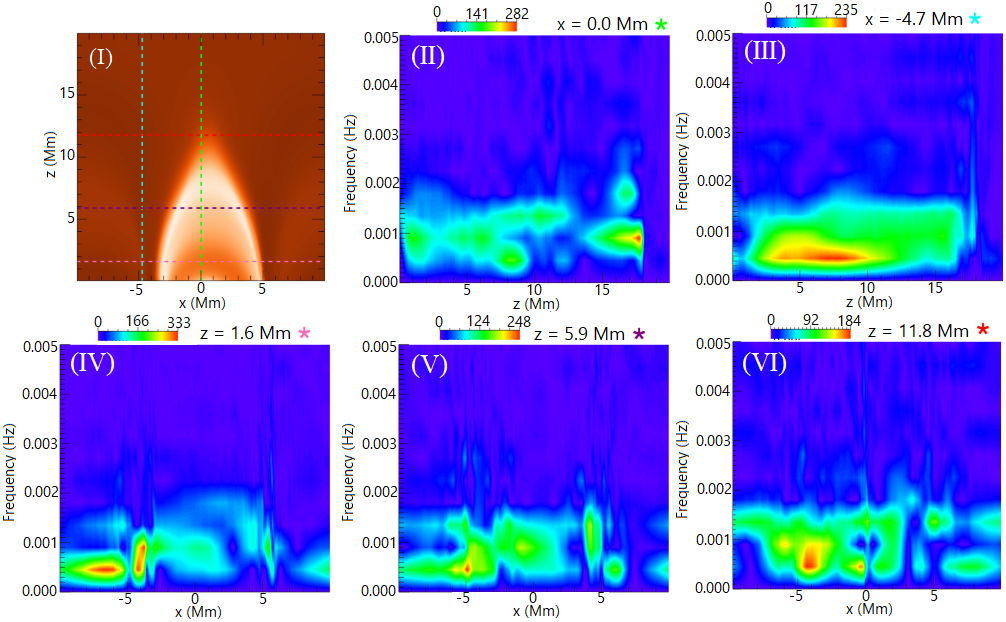}}
  \caption{}
  \label{braid_fft_dv}
\end{subfigure}
\begin{subfigure}{0.8\textwidth}
  \centering
  \hspace{0cm}
\makebox[0pt]{\includegraphics[width=1.\textwidth]{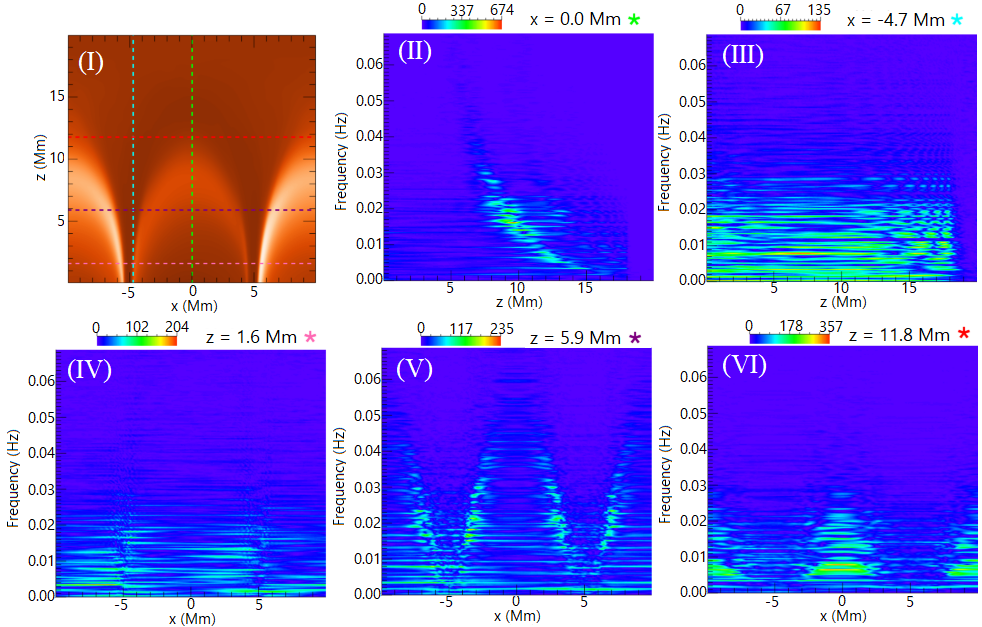}}
  \caption{}
  \label{wave_fft_dv}
\end{subfigure}
\caption{FFT of the Doppler velocity for Fe \rom{12} along various slits in the ideal (a) $\TDC$ and (b) $\TAC$ simulations in panels (\rom{2})-(\rom{6}). The slits are colour coded in the synthetic total intensity figure in panel (\rom{1}).}
\label{fft_dv}
\end{figure*}

\begin{figure}[h!]
\centering
\vspace{0cm}
\begin{subfigure}{0.25\textwidth}
  \centering
  \hspace{0cm}
  \makebox[0pt]{\includegraphics[width=1.\textwidth]{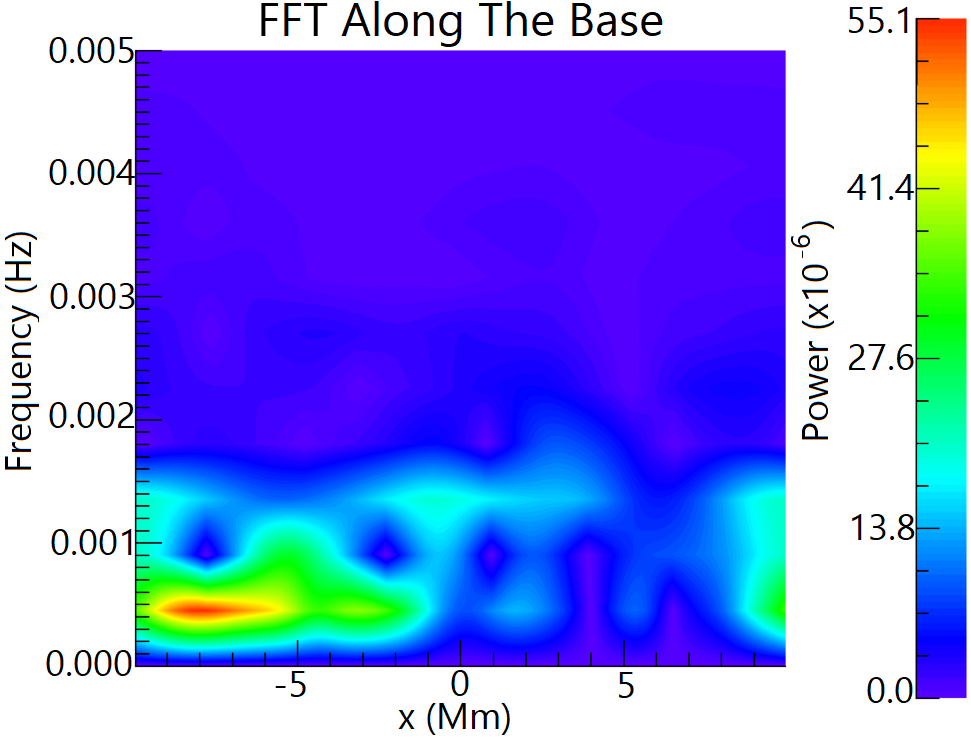}}
  \caption{}
  \label{braid_Driver_vy_fft}
\end{subfigure}%
\begin{subfigure}{0.25\textwidth}
  \centering
  \makebox[0pt]{\includegraphics[width=1.\textwidth]{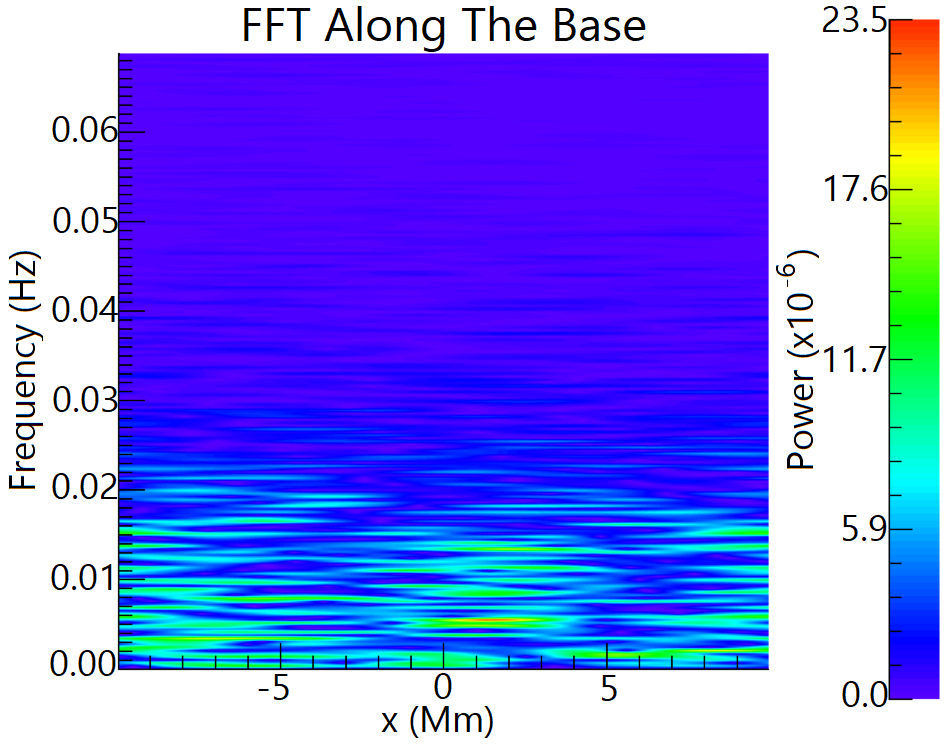}}
  \caption{}
  \label{wave_driver_vy_fft}
\end{subfigure}
\caption{FFT on the bottom $z$ boundary of the LOS average $v_y$ of the drivers (a) $\TDC$ and (b) $\TAC$. The range in the frequency axis changes between each panel.}
\label{fft_of_drivers}
\end{figure}

The behaviour of the lower, more dominant Doppler velocity frequencies, which are the results of the boundary driver, can also be investigated by implementing an FFT. Figure \ref{fft_dv} shows the power of the frequencies observed along the five slits (panels \rom{2}-\rom{6}) across the POS (panel \rom{1}) for the (a) $\TDC$ and (b) $\TAC$ simulations. As expected, the long timescale footpoint motions ($\TDC$) produce low frequency velocity perturbations throughout the domain compared to the short timescale driving ($\TAC$). This is seen by the smaller frequency range given for the $\TDC$ simulation in comparison to the $\TAC$ simulation (see Figs. \ref{braid_fft_dv} and \ref{wave_fft_dv}, respectively). These smaller and larger frequencies, as well as the structure of the FFTs in Fig. \ref{fft_dv}, are explained by the FFT of the $v_{y}$ component of the velocity driver averaged along the LOS for each driver (see Fig. \ref{fft_of_drivers}). 


There are some structural features present in the FFT contours of the $\TAC$ simulation (see Fig. \ref{wave_fft_dv}). Firstly, we shall examine the two vertical slits (panels \rom{2} and \rom{3}). When the slit is positioned between two arcades (panel \rom{3}) it is approximately parallel to the field. As a result the frequencies are similar along the slit. In comparison, when the slit is positioned through the apex of the arcade (panel \rom{2}), the slit passes through a multitude of magnetic field lines, all of varying length. From the FFT it is clear that the frequencies decrease with increasing altitude. This is because longer magnetic field lines (higher altitudes) have lower natural \Alfven frequencies. This relation is also apparent in the triangular features seen through the horizontal slits in panels \rom{5} and \rom{6}. 
 

\begin{figure*}[t!]
\centering
\vspace{0cm}
\begin{subfigure}{0.8\textwidth}
  \centering
  \makebox[0pt]{\includegraphics[width=1.\textwidth]{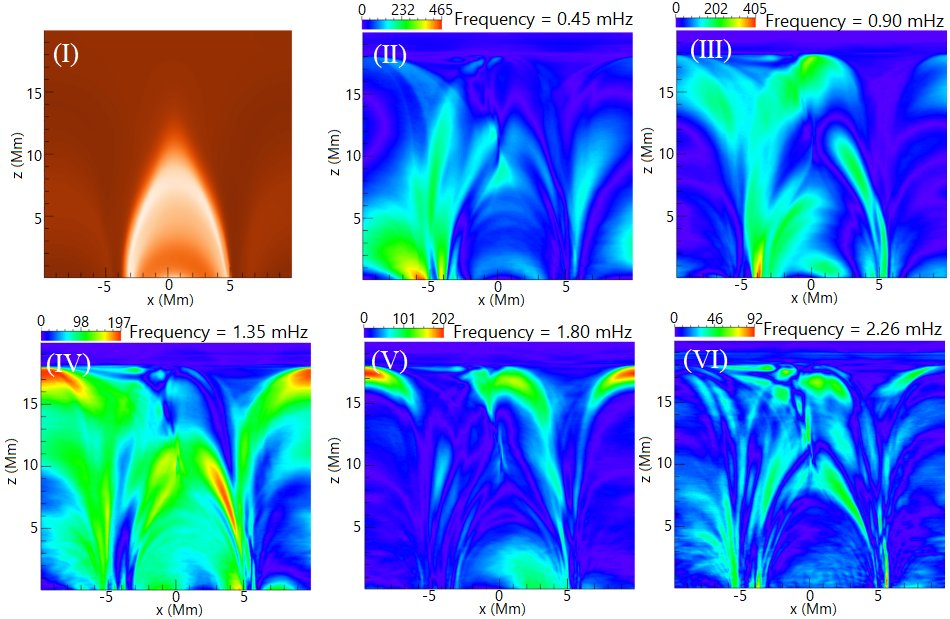}}
  \caption{}
  \label{braid_power_at_specific_freq_dv}
\end{subfigure}
\begin{subfigure}{0.8\textwidth}
  \centering
  \hspace{0cm}
\makebox[0pt]{\includegraphics[width=1.\textwidth]{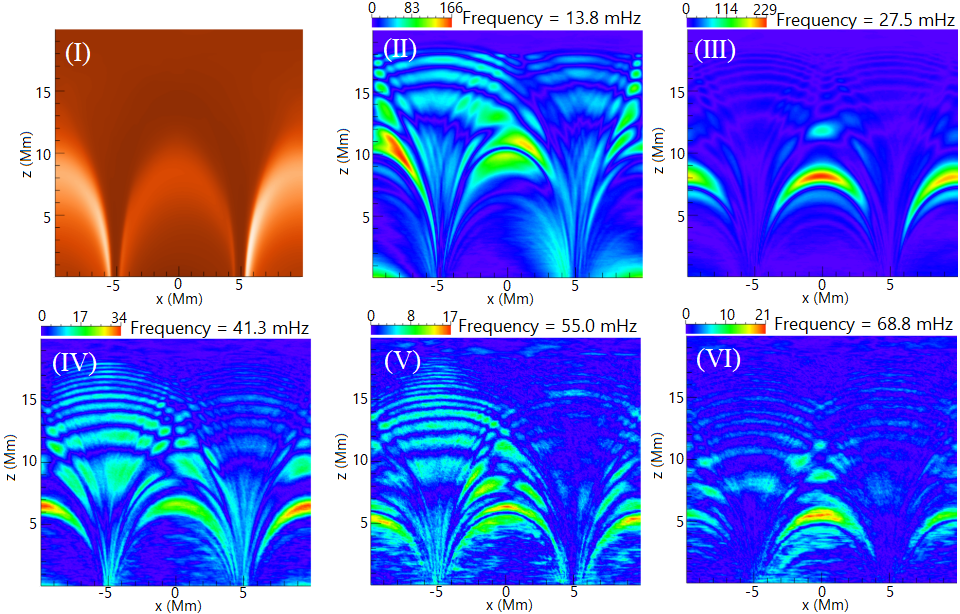}}
  \caption{}
  \label{wave_power_at_specific_freq_fft_dv}
\end{subfigure}
\caption{Power from the FFT of the Fe \rom{12} Doppler velocity at increasing frequencies through panels \rom{2} - \rom{6} in the ideal (a) $\TDC$ and (b) $\TAC$ simulations. For reference, panel \rom{1} illustrates the synthetic total intensity in Fe \rom{12}.}
\label{power_at_specific_freq_fft_dv}
\end{figure*}

Another way to investigate the FFT results is to select a frequency and then analyse the power of this frequency in the POS. Figure \ref{power_at_specific_freq_fft_dv} shows the power in the POS, with increasing frequencies through panels \rom{2} to \rom{6}, from an FFT analysis for the (a) $\TDC$ and (b) $\TAC$ simulations. As the $\TDC$ driver generates lower frequencies than the $\TAC$ driver, different frequencies are used in Figs. \ref{braid_power_at_specific_freq_dv} and \ref{wave_power_at_specific_freq_fft_dv}, respectively. Within the $\TDC$ simulation, there is no clear correlation between the frequencies and their locations. However, it is evident that with increasing frequency the power decreases as the frequencies are greater than those present in the $\TDC$ simulation. Previously, it was shown in Fig. \ref{wave_fft_dv} that with increasing altitude (i.e. increasing loop length) the frequencies present in the FFT would decrease. This is indeed the case here for the $\TAC$ simulation (see Fig. \ref{wave_power_at_specific_freq_fft_dv}). With increasing frequency (panels \rom{2} through \rom{6}), the regions of higher power, which approximately outline with arcade's magnetic field lines, decrease to lower altitudes. 

\begin{figure}[h!]
\centering
\vspace{0cm}
\begin{subfigure}{0.25\textwidth}
  \centering
  \hspace{0cm}
  \makebox[0pt]{\includegraphics[width=1.\textwidth]{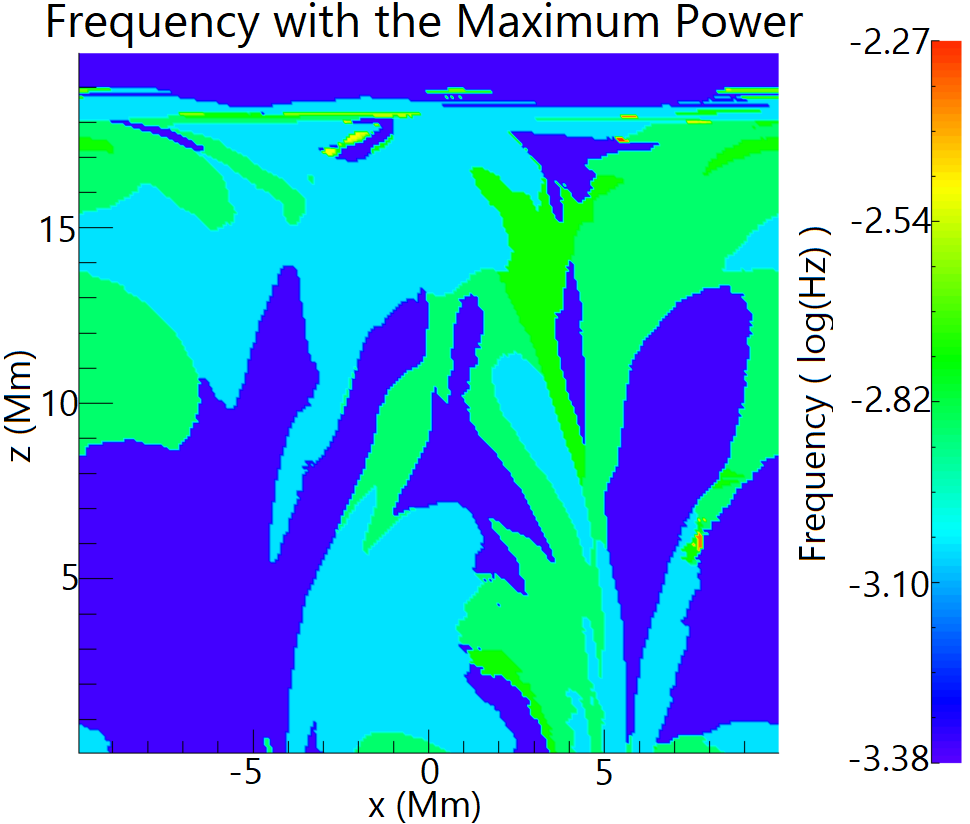}}
  \caption{}
  \label{braid_freq_at_max_power_fft_dv}
\end{subfigure}%
\begin{subfigure}{0.25\textwidth}
  \centering
  \makebox[0pt]{\includegraphics[width=1.\textwidth]{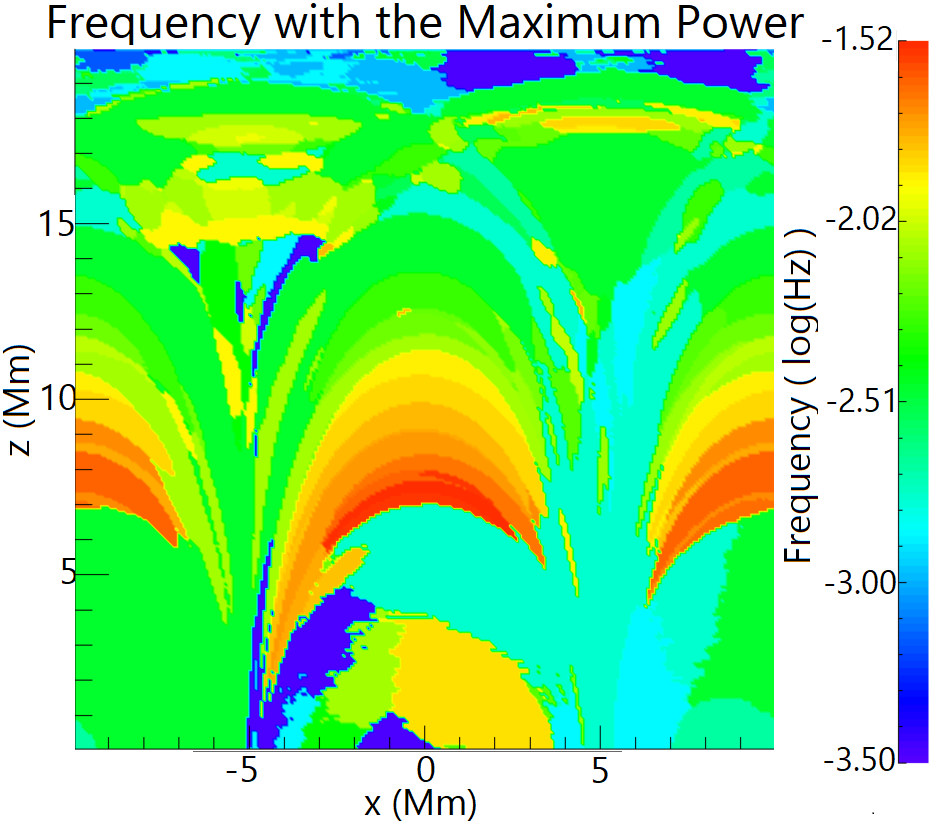}}
  \caption{}
  \label{wave_freq_max_power_fft_dv}
\end{subfigure}
\caption{Logarithm of the frequency with the maximum power of the Fe \rom{12} Doppler velocity FFT in the ideal (a) $\TDC$ and (b) $\TAC$ simulations.}
\label{max_power_fft}
\end{figure}

Our final method of analysing the FFT results produces identical conclusions from the previous FFT analysis. Figure \ref{max_power_fft} shows the frequency with the maximum power of the FFT in the POS for the (a) $\TDC$ and (b) $\TAC$ simulations.  The logarithm of the frequency is taken to help illustrate the features. The $\TDC$ simulation does not show any discernible features whereas the short timescale driving (i.e. $\TAC$ simulation) generates Doppler velocities whose frequency decreases with increasing altitude (i.e. loop length).

\subsection{Line widths}

\begin{figure*}[t!]
\centering
\vspace{0cm}
\begin{subfigure}{0.3\textwidth}
  \centering
  \hspace{0cm}
  \makebox[0pt]{\includegraphics[width=1.\textwidth]{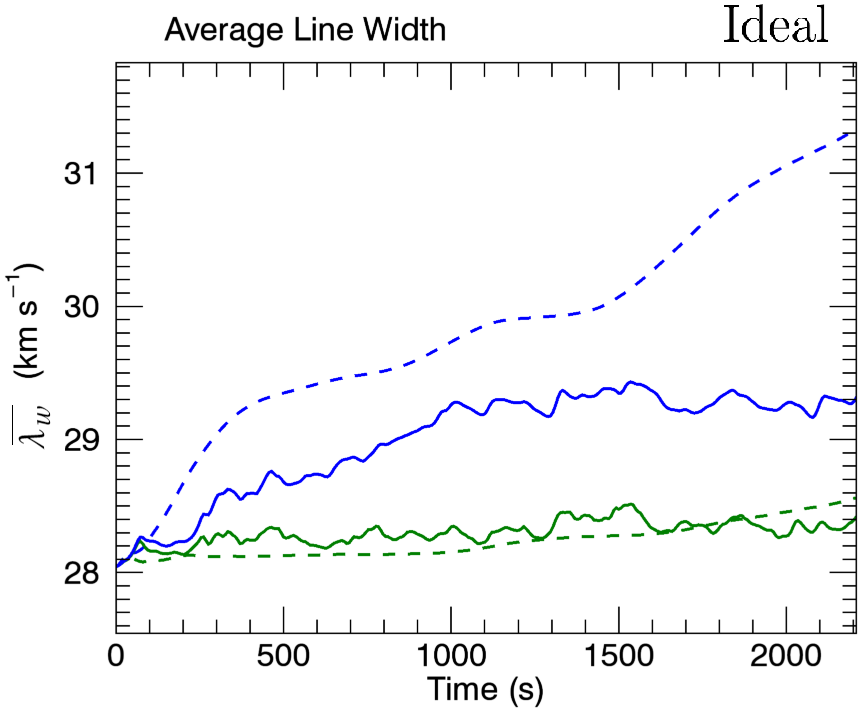}}
  \caption{}
  \label{fwhm_ideal}
\end{subfigure}
\begin{subfigure}{0.3\textwidth}
  \centering
  \makebox[0pt]{\includegraphics[width=1\textwidth]{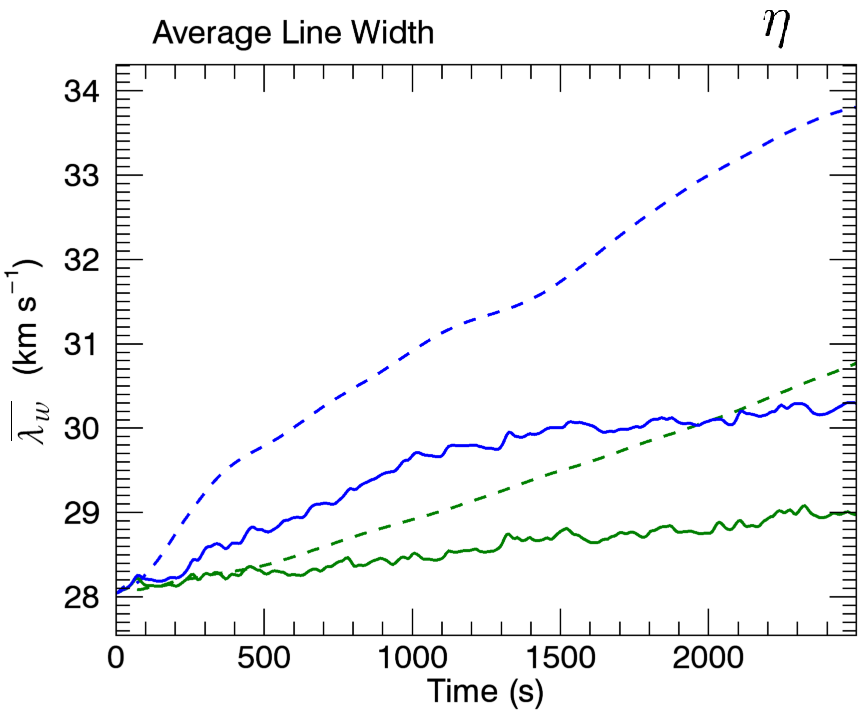}}
  \caption{}
  \label{fwhm_res}
\end{subfigure}
\begin{subfigure}{0.3\textwidth}
  \centering
  \makebox[0pt]{\includegraphics[width=1\textwidth]{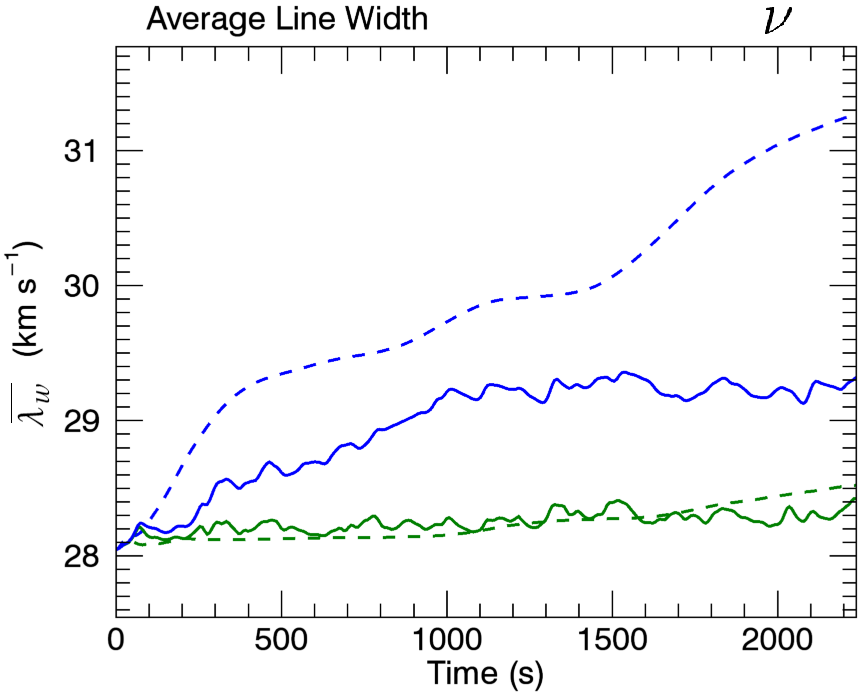}}
  \caption{}
  \label{fwhm_visc}
\end{subfigure}
\caption{The average line width (FWHM), $\bar{\lambda}_{w}$, as a function of time during the $\TDC$ (dashed line) and $\TAC$ (solid line) simulations in the ideal (top panel), resistive (bottom left panel) and viscous (bottom right panel) regimes. Emission lines Fe \rom{9} (green) and Fe \rom{12} (blue) are investigated.}
\label{fwhm}
\end{figure*}

\begin{figure*}[t!]
\centering
\vspace{0cm}
\begin{subfigure}{0.3\textwidth}
  \centering
  \hspace{0cm}
  \makebox[0pt]{\includegraphics[width=1.\textwidth]{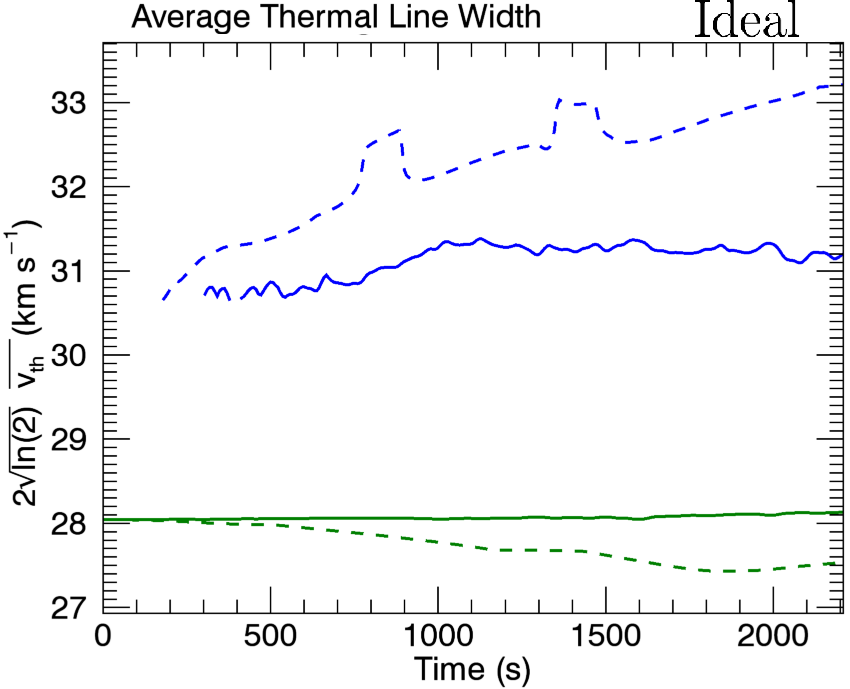}}
  \caption{}
  \label{av_vth_ideal}
\end{subfigure}
\begin{subfigure}{0.3\textwidth}
  \centering
  \makebox[0pt]{\includegraphics[width=1\textwidth]{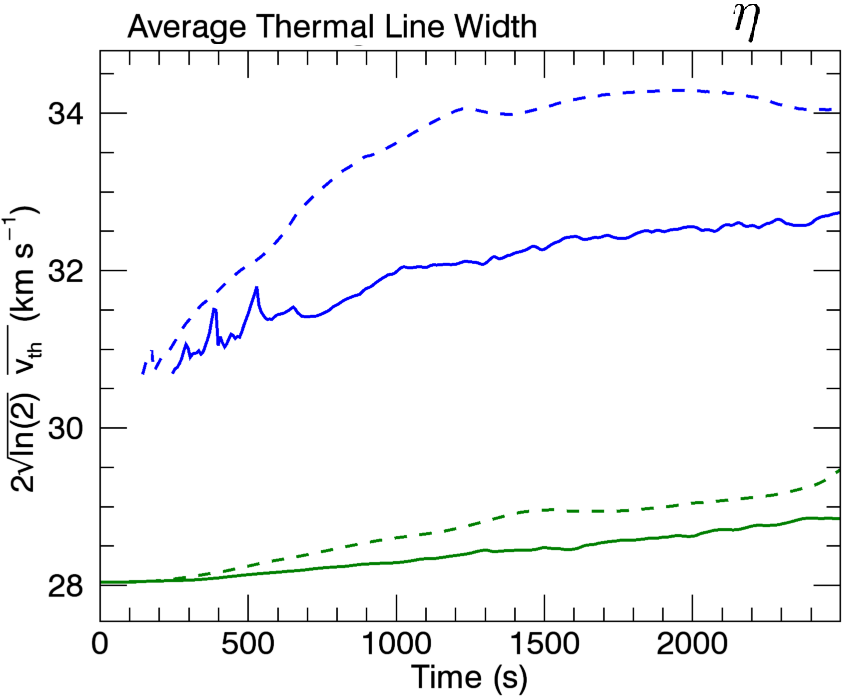}}
  \caption{}
  \label{av_vth_res}
\end{subfigure}
\begin{subfigure}{0.3\textwidth}
  \centering
  \makebox[0pt]{\includegraphics[width=1\textwidth]{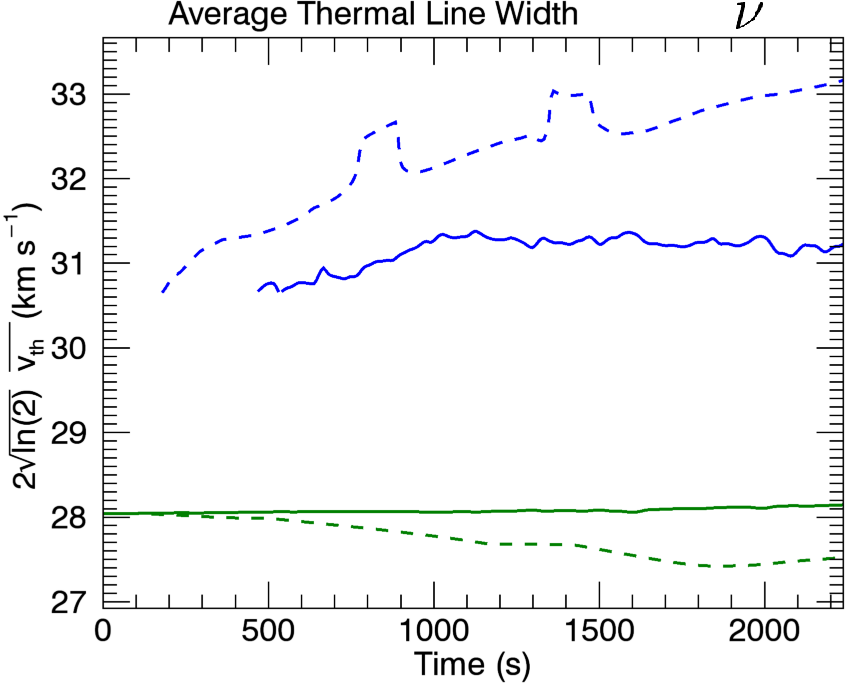}}
  \caption{}
  \label{av_vth_visc}
\end{subfigure}
\caption{The average estimated thermal line width ($\bar{v}_{th}$) as a function of time during the $\TDC$ (dashed line) and $\TAC$ (solid line) simulations in the ideal (top panel), resistive (bottom left panel) and viscous (bottom right panel) regimes. Emission lines Fe \rom{9} (green) and Fe \rom{12} (blue) are examined. Note the multiplication of $2\sqrt{\text{ln}(2)}$ to allow for comparison with Figure \ref{fwhm}.}
\label{estimated_therm_lw}
\end{figure*}

We now investigate the line width of the specific intensity. Figure \ref{fwhm} shows the evolution of the spatial average line width ($\bar{\lambda}_{w}$) calculated using the full-width-half-maximum (FWHM) of the Gaussian fitted to the specific intensity (see Gaussian fitting discussion at the start of Sect. \ref{spec_sig_sect}). Panels (a) - (c) represent the ideal, resistive and viscous simulations for both the $\TDC$ (dashed line) and $\TAC$ (solid line) drivers. We show results for the Fe \rom{9} (green) and Fe \rom{12} (blue) emission lines. Due to the limited differences between the ideal and viscous simulations, the average line widths are very similar (compare panels (a) and (c)). On the other hand, the resistive regime produces line widths of approximately $1-3 \text{ km}\text{ s}^{-1}$ larger than those in the ideal and viscous simulations. The largest increase is observed in the resistive $\TDC$ simulation as a result of the substantial heating. Indeed, this is confirmed by the increase in the spatially averaged estimated thermal line width ($\bar{v}_{\text{th}}$) shown in Fig. \ref{estimated_therm_lw}. Firstly, to estimate the thermal line width, the mean temperature along the LOS ($\bar{T}$) was calculated, excluding points from the calculation with temperatures outside the formation range of the ion (Fe \rom{9}: $\sim$ 0.501-1.33 MK and Fe \rom{12}: $\sim$ 1.15-2.14 MK). Then, the estimated thermal line width ($v_{\text{th}}$) is calculated and then spatially averaged ($\bar{v}_{\text{th}}$). The mean temperature and estimated thermal line width are given by

\begin{eqnarray}
\  \bar{T}_{i, k, l} &=& \frac{1}{n_{y}}\sum_{j = 1}^{n_{y}} T_{i,j,k,l},  \; \\
\ v_{th;i,k,l} &=& \sqrt{\frac{2k_B\bar{T}}{m_{p}\mu_{\lambda_0}}},  \;
\end{eqnarray}

\noindent respectively. The indices of the coordinates $(x,y,z,t)$ are given by $i$, $j$, $k$ and $l$ and $n_y$ denotes the number of points along our LOS (i.e. the $y$ axis). The constants, $k_B$, $m_p$ and $\mu_{\lambda_0}$ denote the Boltzmann constant, the mass of a proton and the atomic weight in proton masses of the emitting element, respectively. The multiplicative constant $2\sqrt{\text{ln}\left(2\right)}$ seen in the estimated thermal line width (Fig. \ref{estimated_therm_lw}) is to allow for a comparison between the thermal line width and the total line width (Fig. \ref{fwhm}). This arises due to the relation between the line width (FWHM) and thermal velocity of a single emission line seen in Eq. 5 of \citet{VanDoorsselaereAntolin2016} (note there is a minor typo: there is an extra $\sqrt{2}$ in their first expression for $\lambda_{w}$). In some instances the estimated thermal line width in Fig. \ref{estimated_therm_lw} is larger than the total line width, which is obviously unphysical. This is due to overestimating the temperature at times within our thermal line width estimation. 

\subsection{Kinetic Energy}

\begin{figure*}[t!]
\centering
\includegraphics[width=.9\textwidth]{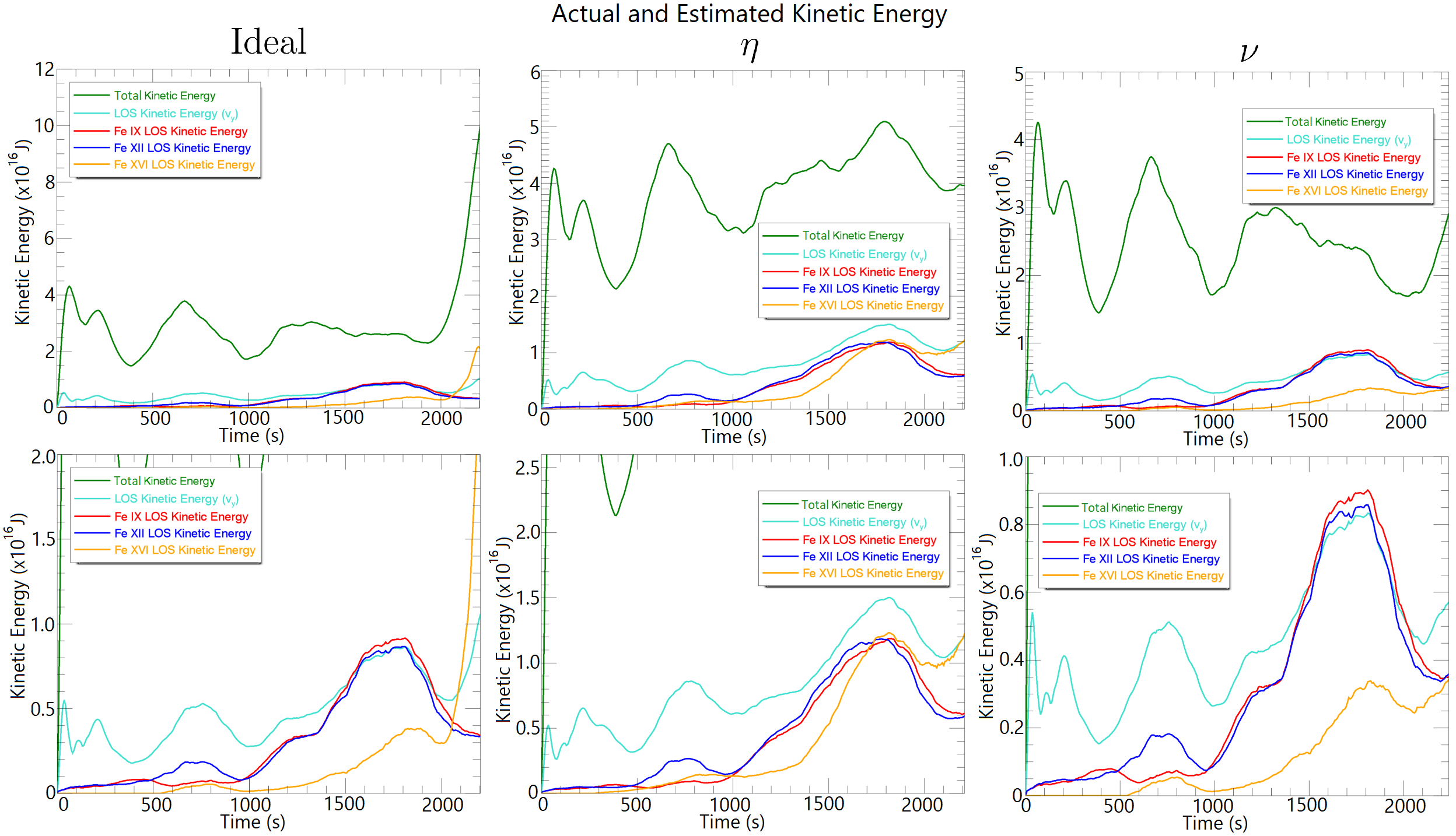}
\caption{Total (green) and LOS (turquoise) kinetic energy integrated over the full 3D numerical domain compared to the estimated kinetic energy in Fe \rom{9} (red), Fe \rom{12} (blue) and Fe \rom{16} (orange), during the ideal (column 1), resistive (column 2) and viscous (column 3) $\TDC$ simulations. The bottom row is the same as the top row but with a reduced kinetic energy range.}
\label{braid_ek}
\end{figure*}

\begin{figure*}[t!]
\centering
\includegraphics[width=.9\textwidth]{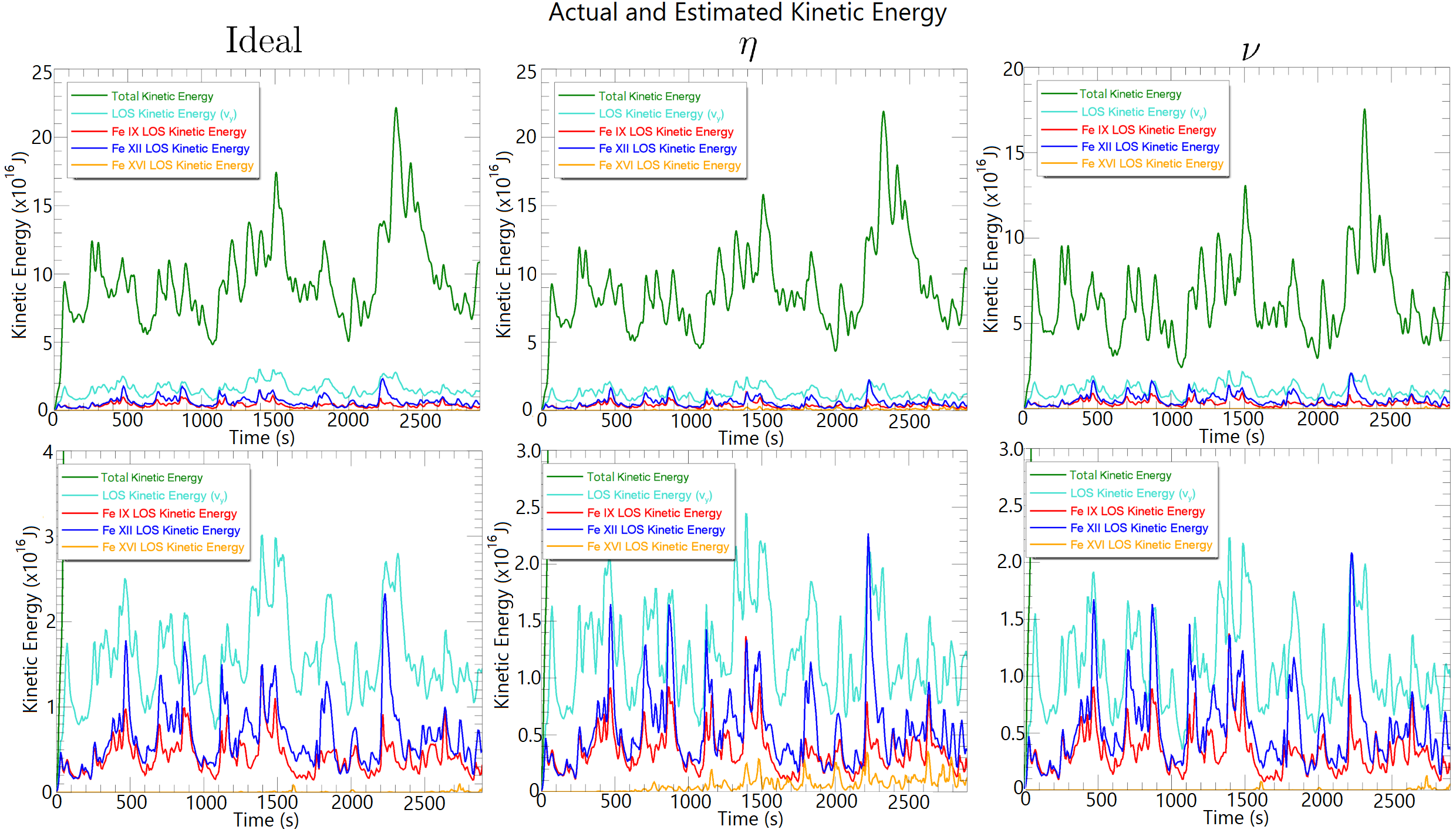}
\caption{Total (green) and LOS (turquoise) kinetic energy integrated over the full 3D numerical domain compared to the estimated kinetic energy in Fe \rom{9} (red), Fe \rom{12} (blue) and Fe \rom{16} (orange), during the ideal (column 1), resistive (column 2) and viscous (column 3) $\TAC$ simulations. The bottom row is the same as the top row but with a reduced kinetic energy range.}
\label{wave_ek}
\end{figure*}

In this section, we use the Doppler velocities, previously analysed in Sect. \ref{spec_sig_sect}, to estimate the kinetic energy, and then compare it to the true kinetic energy in the numerical system. The evolution of the kinetic energies (actual and estimated) using the $\TDC$ and $\TAC$ characteristic timescales are shown in Figs. \ref{braid_ek} and \ref{wave_ek}, respectively. The volume integrated LOS kinetic energy  and the estimated kinetic energy, using the spectroscopic data, are calculated using

\begin{eqnarray}
\;&&\text{LOS Kinetic Energy} = \frac{1}{2}\int\rho v_{LOS}^2 dV, \label{los_ke_eqn} \\
\;&&\text{Estimated Kinetic Energy} = \frac{L}{2}\int\bar{\rho}v_D^2 dA. \label{los_fomo_ke_eqn}  
\end{eqnarray}

\noindent where $v_{\text{LOS}} = v_{y}$, $L$ is the LOS depth, $\bar{\rho}$ is the average density along the LOS, and $v_{D}$ is the Doppler velocity along the LOS (i.e. $y$ axis). Within coronal observations, the LOS depth $L$ and the density profile along the LOS are not easily attainable. The behaviour of the estimated kinetic energy does not change as a result of the length $L$, which is only a scaling factor ($L=20$ Mm). However, to compare the estimated kinetic energy to the actual total kinetic energy, a reasonable estimate of this length is required. As for the density, 
 an average along the LOS can be estimated using line ratios \citep[e.g.][]{DereMason1979, MasonBhatia1979, LandiMiralles2014,PolitoDelZanna2016}.

When comparing the kinetic energy in the $\TDC$ (Fig. \ref{braid_ek}) and $\TAC$ simulations (Fig. \ref{wave_ek}), the first stand out feature is there is more variation in the kinetic energy when a short timescale driver is used \citep{HowsonDeMoortelFyfe2020}. Not only is this seen in the total and LOS kinetic energies but these sharp and frequent gradients are apparent in the estimated kinetic energies too. 

Between the ideal and non-ideal simulations, with the same timescale for the boundary driver, the total kinetic energies do not differ significantly, and the LOS kinetic energies only vary slightly. When comparing the different drivers it is found that the $\TAC$ simulation is approximately just under one order of magnitude greater than the $\TDC$ simulation.

In all the simulations, our chosen LOS (i.e. $y$ axis) generates LOS kinetic energies which underestimate the total kinetic energy. This is to be expected given the multi-directional nature of the velocity driver. As a consequence, even without forward modelling effects, some of the kinetic energy is `lost' along this, and any other, LOS.  Obviously as a result, the estimated kinetic energy underestimates the total kinetic energy. However, it also underestimates the LOS kinetic energy. This is due to both the cancellation of actual velocities along the LOS and detecting plasma which is only within the formation temperature range for the emission line under investigation. A good example of this second case is seen in the Fe \rom{16} estimated kinetic energy. We have shown that Fe \rom{9} and Fe \rom{12} detect more regions of the domain than Fe \rom{16} does, especially in the $\TAC$ simulations. The Doppler velocities at the locations where Fe \rom{16} does not detect the plasma -- due to plasma temperatures outwith the formation temperature of Fe \rom{16} -- are set to zero; hence a larger underestimation of the LOS kinetic energy is obtained in Fig. \ref{wave_ek}. Therefore estimates of the total kinetic energy are even more inaccurate. \citet{HowsonDeMoortelFyfe2020} show that the magnetic energy dominates over the kinetic energy and hence underestimations of the kinetic energy through spectroscopic observations might be insignificant when examining the total energy within the system.


\section{Discussion and conclusions} \label{sec_Discussion}

In this paper, we have examined the synthetic emission data produced from simulations of heating in a coronal arcade. We considered resistive, viscous and ideal simulations for velocity drivers with two different characteristic timescales. The drivers were designed to mimic the motions of magnetic footpoints at the base of the corona, with one case akin to AC motions and DC motions in the other. Within our synthetic emission analysis, we used a LOS parallel to the invariant direction of the initial magnetic field (see Fig. \ref{mag_field_arcade}).

Our model is initially invariant along the $y$ axis, which may not accurately represent the complexity of the corona. As a result, some physical processes (which require a genuinely 3D structure) such as mode coupling and resonant absorption may be inhibited at the start of the simulations. However, as the simulations progress, perturbations caused by the footpoint driving produce genuinely 3D structures \citep[e.g. see Fig. 5 from][]{HowsonDeMoortelFyfe2020} and we do not see any overall change in the relative efficiency of AC and DC drivers once these form. As such, we expect heating to remain larger in the DC cases, if a more general field configuration was investigated. Furthermore, even with a genuinely 3D magnetic field structure, the field strength will remain largest at the magnetic footpoints. Therefore, we expect that Ohmic heating will still dominate at low altitudes. Despite these points, a 3D system will have implications for the interpretation of observables. In this article, we take advantage of a specific LOS which is parallel to the (initial) invariant direction. This means the magnetic topology is identifiable in the synthetic emission. Of course, real observations of a less-ordered coronal arcade will be more difficult to analyse.

Within the total intensity, running difference of the total intensity and the Doppler velocity, the arcade's arch-like structures of the magnetic field are visible regardless of the timescale of the velocity driver. These magnetic field structures are easily identified even though they cannot be measured directly. However, this is not the case when an emission line is chosen which does not fully represent the plasma temperatures present. Indeed, the plasma temperatures within the resistive  AC simulation were smaller than the formation temperature range of Fe \rom{16} and hence only the hotter, low lying magnetic field lines were detected. \citet{Cargill1994} demonstrated, numerically, that plasma exceeding 5 MK is present in loops which have been impulsively heated by nanoflares. The hottest emission line used in this article (Fe \rom{16}) has a peak and maximum formation temperature of approximately 2.65 MK and 6.31 MK, respectively. Therefore, if a hotter emission line is used, such features may be observable. However, in our model we did not include a hotter line as the high temperatures present may be unrealistic due to the absence of thermal conduction and optically thin radiation.

Distinguishing between the associated heating from the AC and DC driving is difficult. The biggest difference we see is the magnitude of the temperature increase (largest for DC driving). However, there is no apparent difference in the spatial distribution of energy release in either case. This translates to the synthetic total intensity, where identifying whether a particular heating event is driven by DC or AC heating may not be possible.

When it comes to comparing Ohmic and viscous heating, resistive simulations do exhibit signatures which are not observed in the viscous cases. For example, in agreement with \citet{VanDoorsselaereAndries2007}, the Ohmic heating is found to occur preferentially close to magnetic foot points. Meanwhile, viscous heating is found to occur preferentially at the separatrix surfaces between the coronal arcades  \citep[see][]{HowsonDeMoortelFyfe2020}. However, as there is relatively little energy in the velocity field, the magnitude of this viscous heating is low. As such, we do not see significant differences between the viscous and ideal simulations. As a result, signatures of viscous heating are less visible than those of Ohmic heating and, in these results, are often obscured by changes in the temperature due to adiabatic effects.

As a result of the short-lived velocity perturbations at the magnetic footpoints, short spatial scales are seen in the running difference of the total intensity and the Doppler velocity within the AC simulation.  These short timescales in the driver cause higher frequencies than those in the DC simulation. This can be detected via a multitude of methods: analysing the Doppler velocity along a slit as a function of time, taking an FFT along this slit, examining the power of an FFT of Doppler velocity over the POS, and investigating the frequency of the maximum power of a Doppler velocity POS FFT. All of these methods were able to conclude that higher frequencies are present in the AC simulation in comparison to the DC simulation. We were able to identify the signatures of standing \Alfvenic waves (see below), despite the random nature of the driver. This resulted in higher frequencies being observed for shorter field lines with higher natural \Alfven frequencies within the $\TAC$ simulation.


Propagating features were found when observing the Doppler velocity along vertical slits. These were interpreted as \Alfvenic waves and dependent on the position of the slit they appeared to propagate at different rates. A slit which passes through the apex of the arcade, and hence crosses multiple magnetic field lines, produces a slower rate of propagation compared to a slit positioned between two arcades. This is because the distance travelled along the curved magnetic field lines, from the base of the corona, to reach a given height on the slit is further than the distance travelled along the magnetic field lines which approximately line up with the slit between the arcades. Several different propagating features are identified when observing the running difference of the total intensity along a vertical slit lined up with the apex of the arcade. Some of these features were observed to propagate with the local \Alfven speed and others with the local fast speed. Therefore, within observations, differentiating between \Alfvenic and fast waves is attainable as well as identifying the speeds. However, in observations signatures may not be as coherent, especially if the LOS is not directed along the invariant direction of the arcade as in our model.

When observations are taken along horizontal slits, a sharp contrast is seen within the local Doppler velocity at the separatrix surfaces when examining a time-distance plot. These features were more apparent in the DC simulation but still present in the AC simulation. Such a structure occurs due to the magnetic connectivity changing across the separatrix surface and hence neighbouring fieldlines may not be excited in the same way \citep{HowsonDeMoortelFyfe2020}. This suggests that Doppler shifts are able to identify regions where the connectivity of the coronal field changes.

As well as observing the low frequency Doppler signals during the evolution of the DC simulation,  higher frequencies with smaller amplitudes are also present. An estimate of the \Alfven travel time was used to approximate the period of the fundamental mode. This was found to reasonably match the dominant frequencies present in the high frequency signal and hence, suggests that these frequencies are the result of an \Alfvenic mode. 

The analysis of the line width showed a larger increase within the resistive DC simulation. This is a result of the increased heating present as the magnetic energy is more readily released. Indeed, this increase was confirmed to be a thermal increase as the thermal line width throughout the full 3D domain was estimated and increased more significantly than the other simulations. 

An obvious comparison between the estimated kinetic energy (using the spectroscopic data) in the AC and DC simulations was the temporal scales. The AC simulation generated more variation in the kinetic energy as a results of the short timescale velocity driver. When it came to predicting the kinetic energy within the simulations, the LOS caused an underestimation of the kinetic energy. This would be the case for any LOS in this simulation as it is the result of the multi-directional velocity driver, meaning that not all the flows present will line up along a chosen LOS. In the simulations presented here, it is reasonable to assume that there is an equipartition between $v_x$ and $v_y$ as there is no preferential direction for the horizontal velocity in the driver. Therefore, we should be able to compensate for some of the `lost' kinetic energy. However, this is not necessarily the case in other models or observations. Two further factors which affect the estimation of the kinetic energy are cancellation of velocity perturbations along the LOS and plasma temperatures which sit outwith the formation temperature of the selected emission line. Improvements to the former can be achieved by examining the non-thermal line widths which can account for the discrepancy between the observed wave energy through Doppler velocities and the true energy in the system \cite[e.g.][]{DeMoortelPascoe2012,McIntoshDePontieu2012,PantMagyar2019}. As for the latter, using DEM analysis may minimise the underestimation of the kinetic energy by allowing the most appropriate emission line to be used. Therefore, estimates of kinetic energies within such observations may not capture a significant proportion of the true energy but methods can be used to minimise the error. \citet{HowsonDeMoortelFyfe2020} showed that the magnetic energy in the simulation dominates the kinetic energy and hence underestimating the kinetic energy does not significantly impact estimates of the total energy within these simulations. However, accurately measuring the observed coronal magnetic field strength and, hence, the associated magnetic field strength is still challenging. Estimates of the kinetic energy are still important in models where the kinetic energy is expected to be comparable to the free magnetic energy, for example in wave models.

In summary, the majority if not all of the observables we have investigated in this paper are relatively easily understood with the knowledge of the evolutionary behaviour of the plasma parameters from the underlying numerical model. However, within observations these parameters are not well known. Our main goal was to establish signatures which could be identified in real observations to distinguish between AC and DC-like footpoint motions. Firstly, we are able to identify the magnetic structure of the arcade, where the presence of separatrix surfaces is revealed directly by the magnetic field structure but also by contrasts in the local Doppler velocity on either side of the connectivity change. The clearest distinction between the AC and DC models are the frequencies detected through analysis of the Doppler velocities where, as expected, the DC model exhibits lower frequencies. This is also observed in the spatial scales of the POS Doppler shifts and running difference of the total intensity. Estimating the kinetic energy within the system is challenging because various factors (e.g. multi-directional driver and velocity cancellation along LOS) prohibit an accurate estimate. Lastly, we note that our model has several limitations (no thermal conduction or optically thin radiation and no coupling to the lower atmosphere) which are likely to affect some of the observational signatures. For example, the extremely large temperatures present in our resistive models will decrease if radiation and thermal conduction are present and this will alter the observables. However, we focused on emission lines with lower temperatures to avoid analysing the unrealistically high temperatures. In addition, we do not expect a large effect on the flows and hence, the analysis of the Doppler velocities presented in this paper.


\vspace{2cm}



\vspace{1cm}
{\emph{Acknowledgements.}} The research leading to these results has received funding from the UK Science and Technology Facilities Council (consolidated grants ST/N000609/1 and ST/S000402/1) and the European Union Horizon 2020 research and innovation programme (grant agreement No. 647214). IDM acknowledges support from the Research Council of Norway through its Centres of Excellence scheme, project number 262622.

\bibliographystyle{aa}        
\bibliography{arcade_paper.bib}           

\end{document}